\documentclass[notitlepage, superscriptaddress, 12pt, one-column, double-spaced]{revtex4-1}
\usepackage{graphicx}
\usepackage{subfigure}
\usepackage{amsmath}

\begin{document}

\title{Resilience of modular complex networks} 

\author{Saray Shai}
\affiliation{School of Computer Science, University of St Andrews, St Andrews, Fife KY16 9SX, Scotland, UK}
\author{Dror Y. Kenett}
\affiliation{Center for Polymer Studies and Department of Physics, Boston University, Boston, Massachusetts 02215, USA}
\author{Yoed N. Kenett}
\affiliation{Gonda Brain Research Center, Bar-Ilan University, Ramat-Gan, Israel}
\author{Miriam Faust}
\affiliation{Gonda Brain Research Center, Bar-Ilan University, Ramat-Gan, Israel}
\affiliation{Department of Psychology, Bar-Ilan University, Ramat-Gan, Israel}
\author{Simon Dobson}
\affiliation{School of Computer Science, University of St Andrews, St Andrews, Fife KY16 9SX, Scotland, UK}
\author{Shlomo Havlin} 
\affiliation{Department of Physics, Bar-Ilan University, 52900 Ramat-Gan, Israel}

\date{\today}

\maketitle

{ \bf
Complex networks often have a modular structure, where a number of tightly-connected groups of nodes (modules) have relatively few interconnections. Modularity had been shown to have an important effect on the evolution and stability of biological networks~\cite{BulBS12}, on
the scalability and efficiency of large-scale infrastructure~\cite{EriESMS03, GuiGMTA05}, and the development of economic and social systems ~\cite{GarGAH08,LuiLW07}. An analytical framework for understanding modularity and its effects on network vulnerability is still missing. Through recent advances in the understanding of multilayer networks~\cite{buldyrev2010catastrophic,GaoGBSH12,LeiLD09}, however, it is now possible to develop a theoretical framework to systematically study this critical issue. Here we study, analytically and numerically,  the resilience of modular networks under attacks on interconnected nodes, which exhibit high betweenness values~\cite{HolHKYH02,SchSMAHH11} and are often more exposed to failure~\cite{BasBBKHI12,GuiGMTA05,MeuMAMB09,ShiSYGLGS12}. Our model provides new understandings into the feedback between structure and function in real world systems, and consequently has important implications as diverse as developing efficient immunization strategies, designing robust large-scale infrastructure, and understanding brain function. 
}

\clearpage 

Network science has become a leading approach to the study of emergent collective phenomena in complex systems, with a wide range of applications to fundamental real word systems~\cite{HavHKBBCHKKKKPS12}.
Many real world systems have been shown to exhibit a modular structure, in which smaller clusters of nodes are connected more to each other than to the network at large, which is key to their behavior and functioning~\cite{NewN03mixing,NewNG04,WuWGS06,GleG08}.
For example, recent studies of biological networks show that the deletion of nodes connecting between modules can have a deleterious effect on the network integrity~\cite{HanHBHGBZDWCRV04}, efficiency~\cite{OlaOHK07}, and stability~\cite{YonYWWCYYTZGZE09}. 
Here we provide an analytical framework for studying the robustness of modular networks in the presence of attacks on interconnected nodes.
We study a percolation process on networks consisting of a varying number of modules, $m$, and a varying number of interconnected nodes.
The analytical solution reveals two percolation regimes separated by a critical number of modules $m^*$: for $m<m^*$ one needs to remove all interconnected nodes to break the system, while the modules are almost unaffected internally.
In contrast, for $m>m^*$ one needs to remove only a fraction of the interconnected nodes, before the system collapses. This is due to the fact that for $m>m^*$  the number of interconnected nodes is high and partial removal of these breaks the modules internally, which helps to bring about the rapid collapse of the whole system. 
Our approach can also be used to study analytically attacks on high betweenness centrality nodes, which in modular structures, correspond to interconnected nodes. Such attacks, which have only been studied numerically so far, are considered to be among the most harmful attack strategies~\cite{HolHKYH02,SchSMAHH11}.

\begin{figure}[h]
\begin{center}
\subfigure{\includegraphics[scale=0.212]{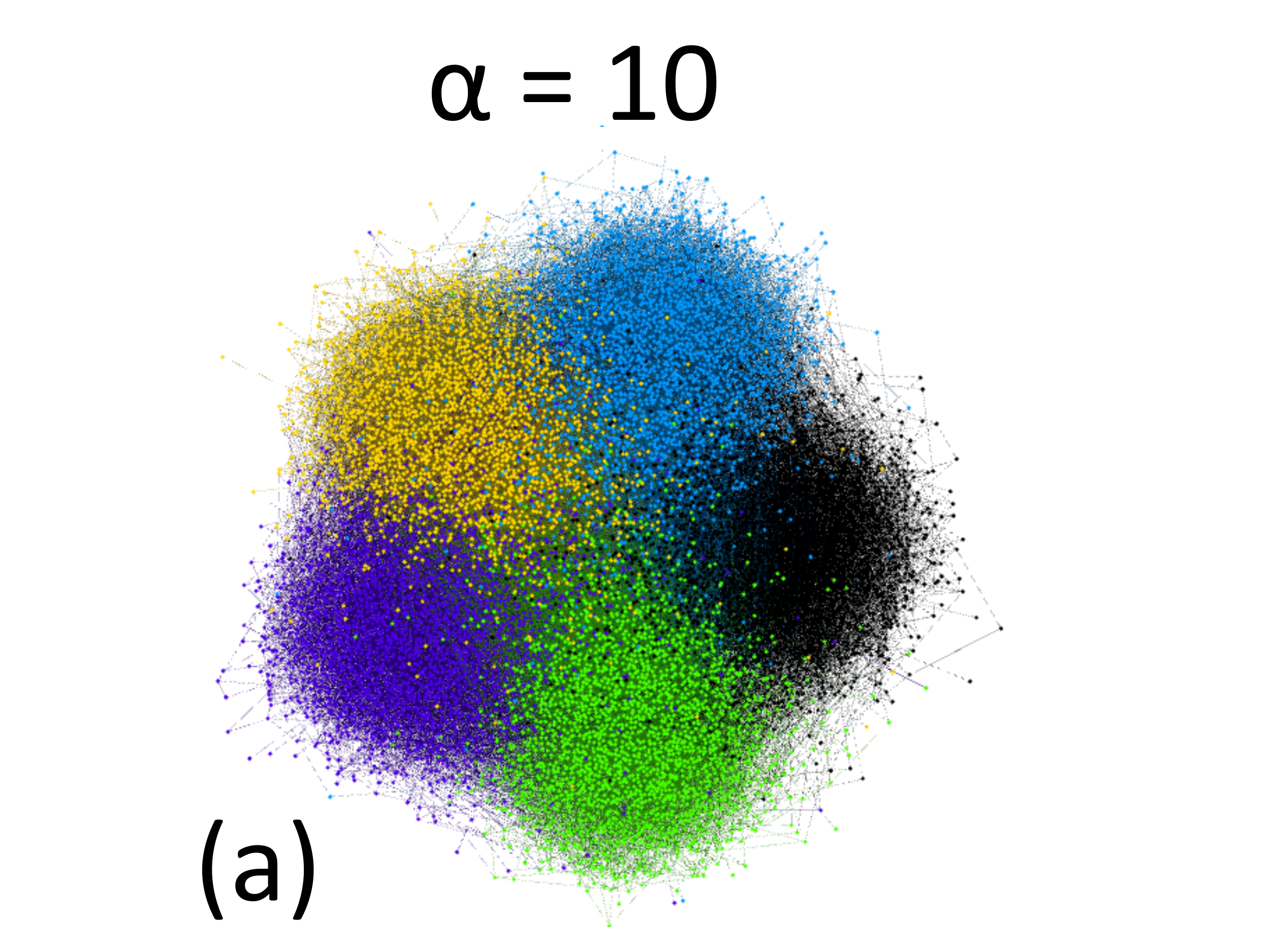}}
\subfigure{\includegraphics[scale=0.212]{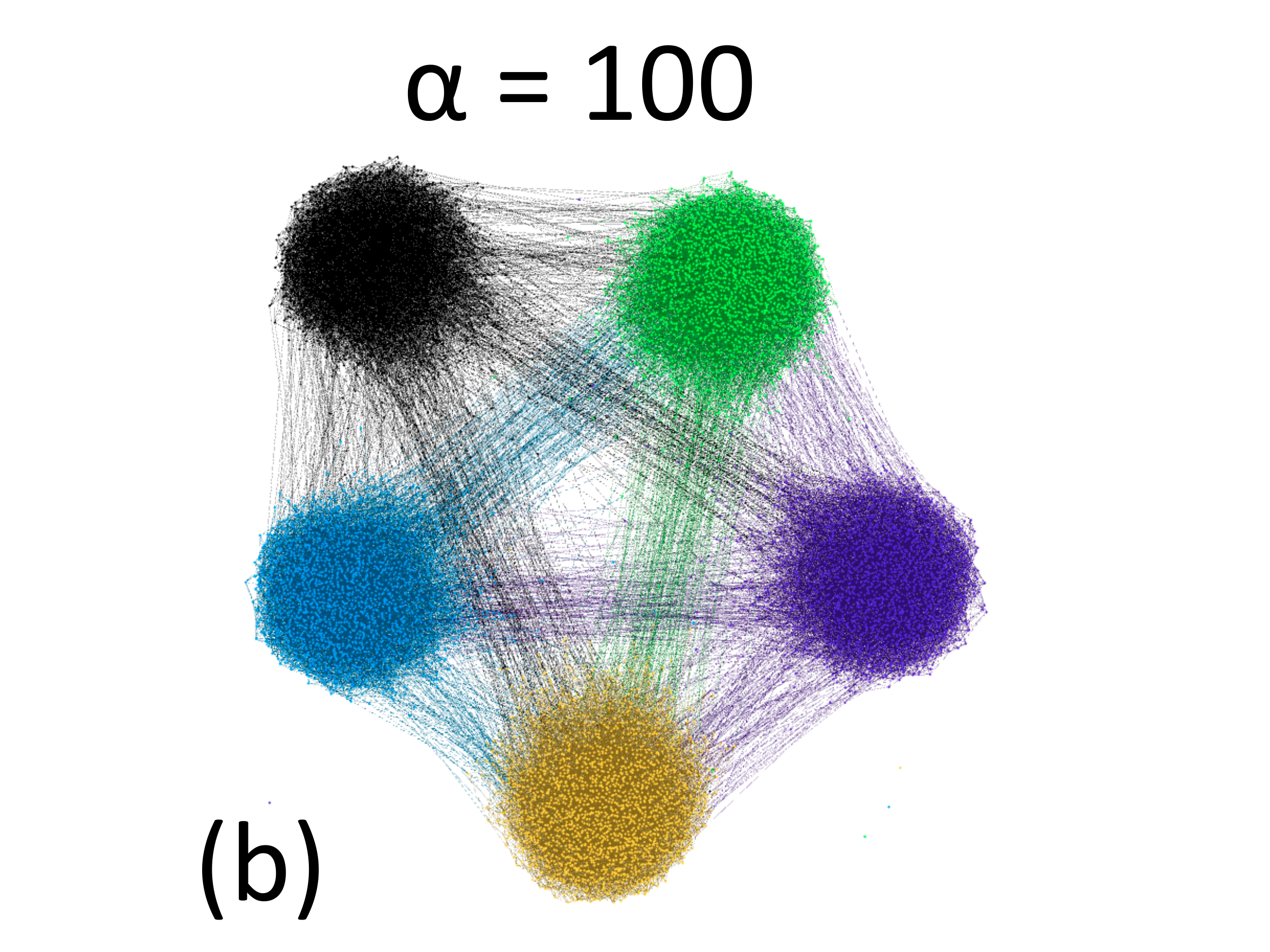}}
\subfigure{\includegraphics[scale=0.212]{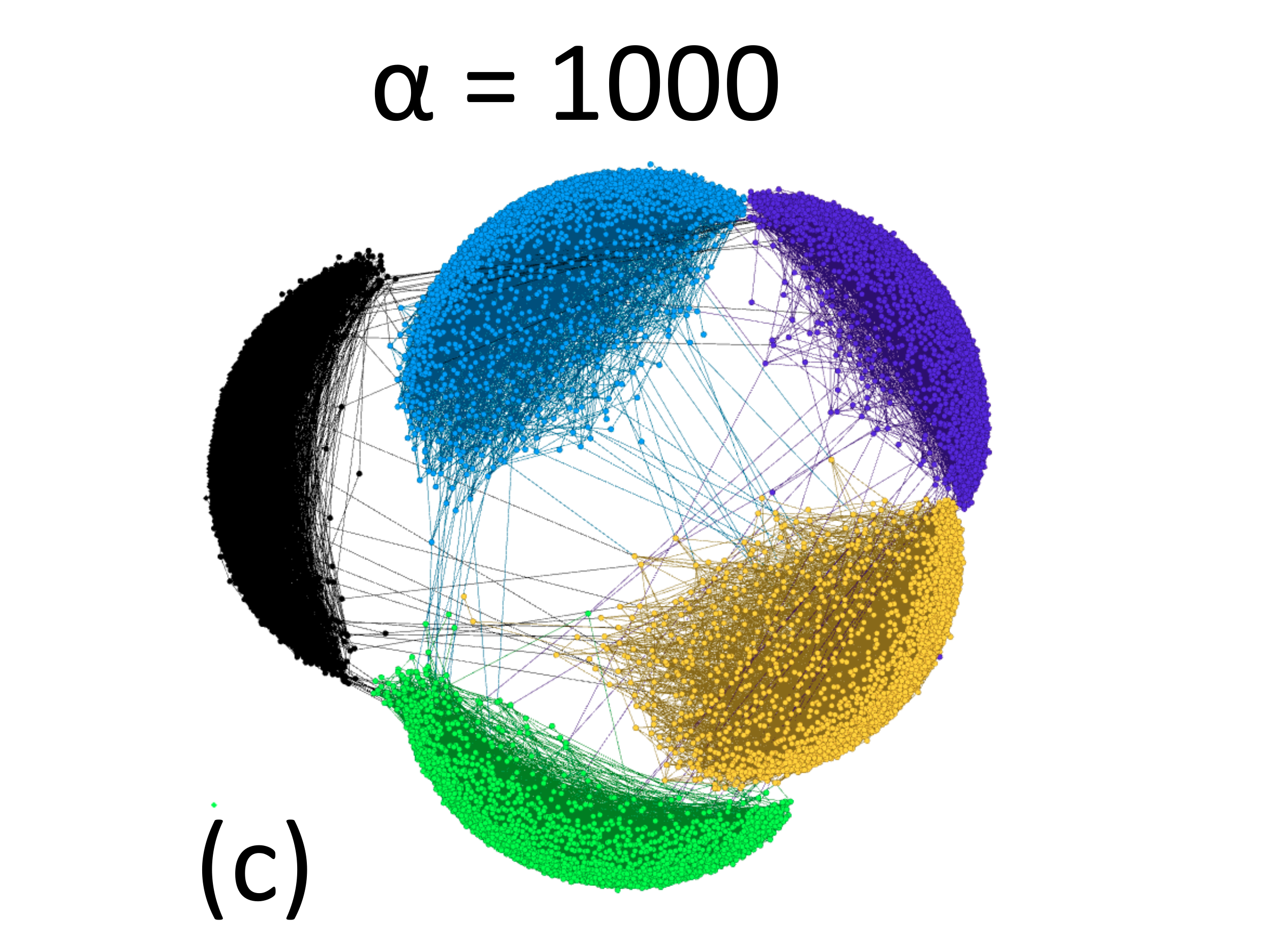}}
\subfigure{\includegraphics[scale=0.38]{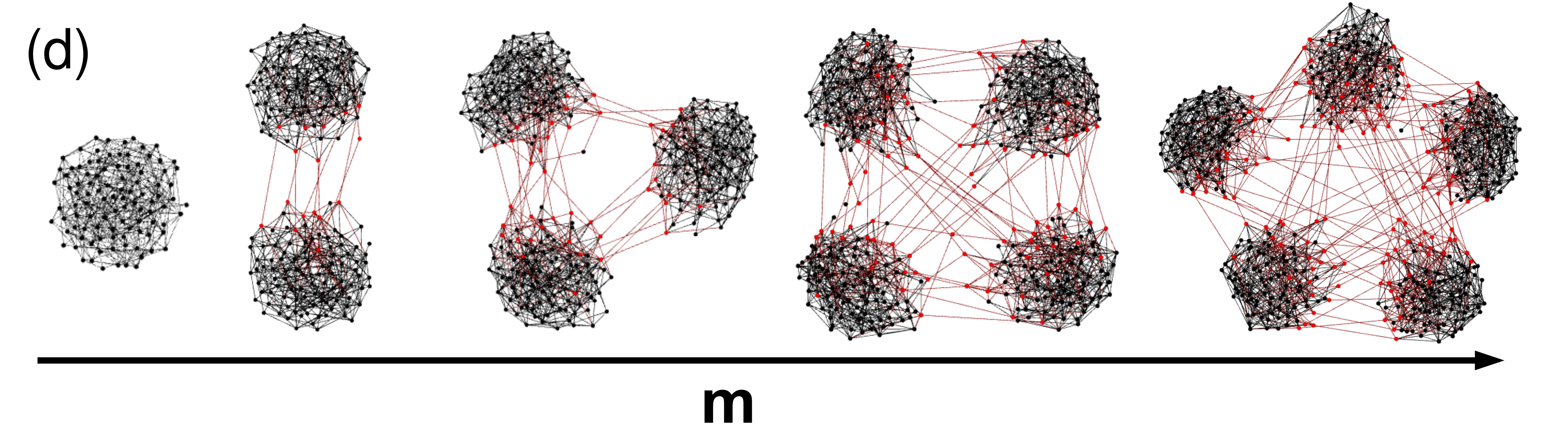}}
\caption{\label{fig:model_vis} {\bf Visualization of the model for generating random modular networks.} (a)-(c) Illustration of the effect of $\alpha$ on the obtained modular network using Gephi with force atlas layout~\cite{BasBHJ09,NoaN09}, on a network of size $N=10\thinspace000$ with mean degree $k=8$ divided into $m=5$ modules. (d) Illustration of the effect of the number of modules $m$ on the obtained network with a number of inter-module links increasing with the number of modules. Inter-connected nodes are shown in red.}
\end{center}
\end{figure}

We consider a modular network with $N$ nodes divided into $m$ equal sized modules.
Similarly to~\cite{NewNG04}, we define $p_{\text{in}}$ as the probability to connect nodes in the same module and $p_{\text{out}}$ as the probability to connect nodes in different modules.
Thus, the total number of intra-module (inter-module) links is given by the probability for a link $p_{\text{in}}$ ($p_{\text{out}}$) multiplied by the number of possible links yielding 
\begin{equation}
M_{\text{in}} = p_{\text{in}} \frac{N (\frac{N}{m} - 1)}{2},
\label{min}
\end{equation}
\begin{equation}
M_{\text{out}} = p_{\text{out}} \frac{N (m-1) \frac{N}{m}}{2}.
\label{mout}
\end{equation}
We define $\alpha$ to be the ratio between the probabilities for an intra- and inter-module link
\begin{equation}
\alpha = \frac{p_{\text{in}}}{p_{\text{out}}}.
\label{alpha}
\end{equation}
In Fig.~\ref{fig:model_vis}(a)-(c) we present an example of modular networks generated with different values of $\alpha$, and visualized using force-directed layout, which has been shown to demonstrate network modularity~\cite{NoaN09} .
Note that the ratio between the number of inter-modules links and intra-module links depends not only on $\alpha$, but also on the number of modules
\begin{equation}
\frac{M_{\text{in}}}{M_{\text{out}}} = \frac{p_{\text{in}} (\frac{N}{m} - 1)}{p_{\text{out}} (m-1) \frac{N}{m}} \sim  \frac{p_{\text{in}}}{p_{\text{out}} (m - 1)} = \frac{\alpha}{m-1}.
\label{eq:min_mout_ratio}
\end{equation}
Thus, our model is taking into consideration that systems comprised of more modules have more inter-links, as illustrated in Fig.~\ref{fig:model_vis}(d). See also Supplementary Fig.~S1, where we show the increase of the mean inter-degree as a function of $m$.

Given the model described above for generating random modular networks, we proceed to study percolation properties for such networks.
We consider a modular Erd\H{o}s-R\'{e}nyi (ER) network~\cite{ErdER60,BolB01} where both the intra- and inter-connectivity are Poisson distributed with means $k_{\text{intra}}$ and $k_{\text{inter}}$ respectively. 
Using the generating function approach presented in~\cite{LeiLD09} (see full details in the Supplementary Information), we find that in the presence of random nodes failure, the giant component emerges when the following is satisfied
\begin{equation}
(1 - k_{\text{intra}})(1 - k_{\text{intra}} - \frac{(m-2)k_{\text{inter}}}{m-1}) - \frac{{k_{\text{inter}}}^2}{m-1} = 0.
\label{eq:general_condition}
\end{equation}
This condition yields $k = k_{\text{intra}} + k_{\text{inter}} = 1$ for every $m$, recovering the standard result for single networks without communities. Thus, in the case of random node failure the percolation threshold only depends on the mean degree, $k$.

However, in real systems the interconnected nodes are often more exposed to failure than other nodes.
For example, it has been shown that aging and schizophrenia could result in damage to the interconnected nodes in brain networks~\cite{ShiSYGLGS12,MeuMAMB09}.
In addition, it is often the case that interconnected nodes are considered to be important; for example, the New York City and London airports, which provide an attractive target for attacks~\cite{GuiGMTA05}.
Therefore, in the following, we consider an attack on modular ER networks where the interconnected nodes are randomly removed.
Let ${r_i} ({k_1, k_2, \ldots, k_m})$ be the occupation probability of a node from module $i$ with $k_1$ links in module $1$, $k_2$ links in module $2$ and etc. When the interconnected nodes are randomly removed, this probability is given by
\begin{equation}
{r_i} ({k_1, k_2, \ldots, k_m}) =
\begin{cases}
1,& \text{if }  \sum\limits_{\substack{j=1 \\ j \neq i}}^m k_j = 0\\
q, & \text{otherwise}
\end{cases},
\label{eq:occ_prob}
\end{equation}
where $q$ is the probability that a randomly chosen interconnected node is occupied.
Let $p$ be the general occupation probability, i.e. the probability that a randomly chosen node is occupied.
Since the probability for a node to be interconnected is $1 - e^{-k_{\text{inter}}}$,  i.e. one minus a Poisson distribution with mean $k_\text{inter}$ at $0$, we obtain
\begin{equation}
q = \frac{p - e^{-k_{\text{inter}}}}{1 - e^{-k_{\text{inter}}}}.
\label{eq:occ_general}
\end{equation}

We extend Callaway \textit{et al.}'s approach~\cite{CalCNSW00} for studying the robustness of networks to intentional attacks, from single-module networks to modular networks in a similar approach as in~\cite{LeiLD09} (see full details in the Methods section), and solve for the occupation probability given in~(\ref{eq:occ_prob}), obtaining two possible solutions for the critical occupation probability of interconnected nodes
\begin{align}
& q_c = 0 \label{eq:qc_1} \\
& q_c =  \frac{-b + \sqrt{b^2 - 4ac}}{2a}  \label{eq:qc_2} \\ \nonumber
& \text{where} \ \ \ \ \ \ a = k_{\text{intra}} k_{\text{inter}} e^{-k_{\text{inter}}} \\ \nonumber
& \ \ \ \ \ \ \ \ \ \ \ \ \ \ b = k_{\text{intra}} + k_{\text{inter}} - k_{\text{intra}}e^{-k_{\text{inter}}} - k_{\text{intra}}k_{\text{inter}}e^{-k_{\text{inter}}} \\ \nonumber
& \ \ \ \ \ \ \ \ \ \ \ \ \ \ c = k_{\text{intra}} e^{-k_{\text{inter}}} - 1. \nonumber 
\end{align}
From these solutions, we obtain the critical occupation probability $p_c$, using Eq.~(\ref{eq:occ_general}). 

Due to symmetry, once the giant component emerges ($p > p_c$) the fraction of nodes of module $i$ in the giant component equals to the total fraction of nodes in the giant component, $S_i = S$, and one obtains,
\begin{equation}
S = e^{-k_{\text{inter}}}(1-q) (1 - e^{{-k_{\text{intra}}} S }) + q(1 - e^{-({k_{\text{intra}}} S +  k_{\text{inter}} S)}).
\label{eq:s_i}
\end{equation}
For $k_{\text{intra}} = 0$, only a fraction $q$ of the nodes in the network are connected, and one obtains  $S = q (1 - e^{-k S})$, recovering the standard result for percolation in single networks~\cite{ErdER60,BolB01}.

\begin{figure}[h]
\begin{center}
\subfigure{\includegraphics[scale=0.7]{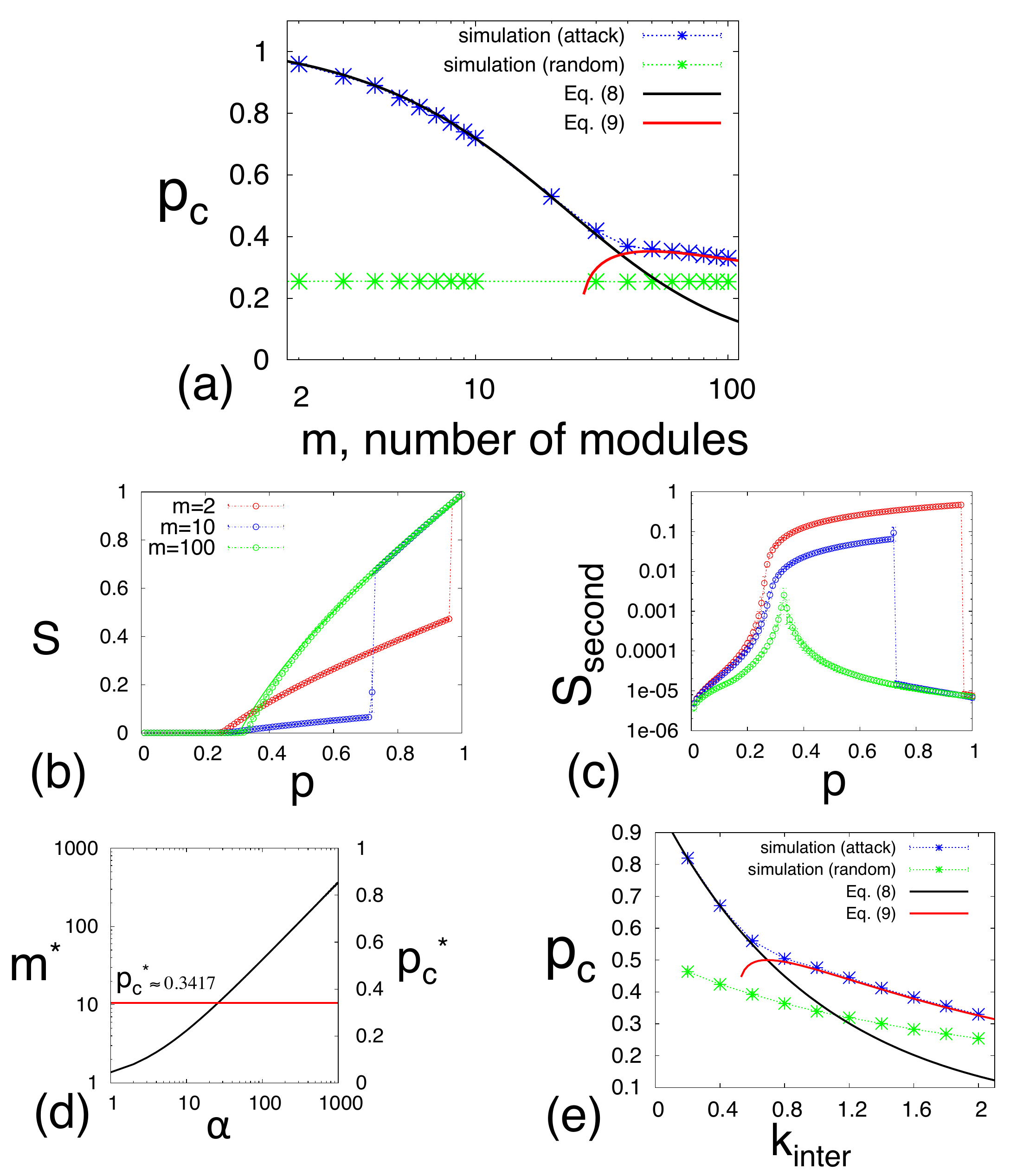}}
\caption{\label{fig:inter_connected_failure} {\bf Two percolation regimes when attacking interconnected nodes.} (a) $p_c$ as a function of $m$ calculated for networks with $k=4$, $\alpha=100$. Simulation points obtained from at least 1000 simulation runs of networks of size $N=600\thinspace000$. Solid lines represent the analytical result obtained in (\ref{eq:qc_1})-(\ref{eq:qc_2}). (b)-(c) Fraction of nodes in the largest cluster $S$ and second largest cluster $S_\text{second}$ as a function of occupation probability $p$. Solid lines represent the analytical result obtained in (\ref{eq:s_i}). (d) Critical number of modules $m^*$, defined as the point where the solutions from (\ref{eq:qc_1}) and (\ref{eq:qc_2}) cross each other, as a function of $\alpha$. ${p_c}^*$ is the percolation threshold at this point. (e) $p_c$ as a function of $k_\text{inter}$ calculated for networks with $m=10$, $k_\text{intra} = 2$.}
\end{center}
\end{figure}

In Fig.~\ref{fig:inter_connected_failure}, we confirm our analytical solution (Eqs.~(\ref{eq:qc_1})-(\ref{eq:s_i})) by extensive numerical simulation of ER modular networks of size $N=600\thinspace000$. 
First, we show the percolation threshold as a function of the number of modules $m$ where the mean degree is kept fixed $k=4$ and $\alpha=100$, see Fig.~\ref{fig:inter_connected_failure}(a). Similar results are shown for $\alpha=10$, and $1000$ in Supplementary Fig.~S2.
Let $m^*$ be the transition point where the two analytical solutions cross each other.
In the regime where $m < m^*$ the attack on interconnected nodes mainly breaks the connectivity between the modules leaving their internal structure intact. Thus, only the removal of all the interconnected nodes ($q_c = 0$) breaks down the giant component.

\begin{figure}[h]
\begin{center}
\subfigure{\includegraphics[scale=0.5]{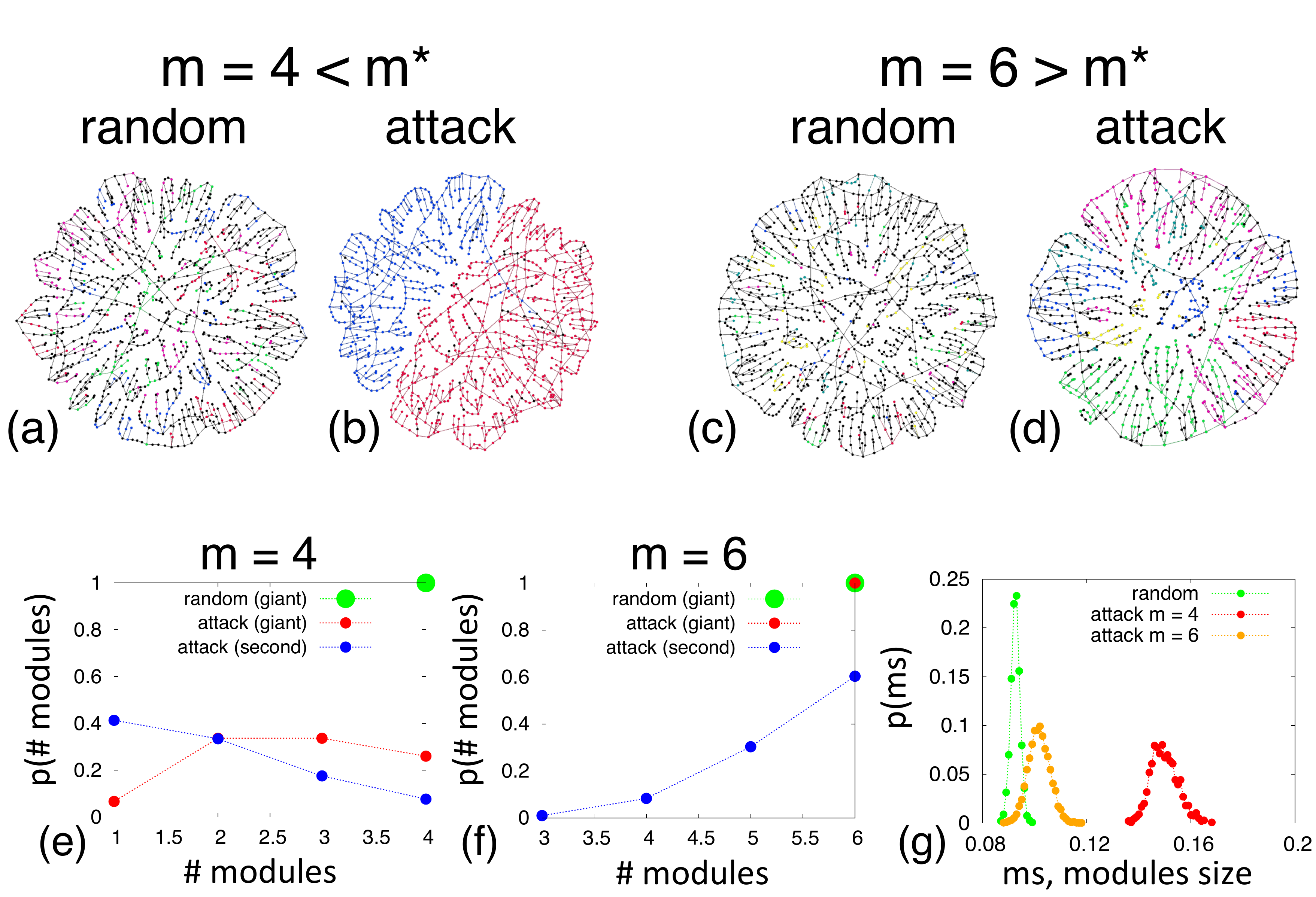}}
\caption{\label{fig:modules_at_s} {\bf Size of modules in the giant component at \boldmath{$S=0.1$}.} Visualization is shown for networks of size $N=12\thinspace000$ with mean degree $k=4$ and $\alpha=10$, at the point where the giant component contains 10\% of the nodes ($S=0.1$) (a),(c) for random node removal, (b),(d) for attack on interconnected nodes. (e)-(f) Distribution of the number of modules in the giant component and second largest component at $S=0.1$. A module is considered to be part of a component if at least one of its nodes are part of the component. (g) Distribution of the size of modules in the giant component at $S=0.1$, normalized by the initial module size. Note that in (g), the size of modules is measured by reconstructing the graph of each module in the giant component, and counting its number of nodes in this graph. In other words, interconnected nodes that have been detached from their original module are not considered. Results obtained by at least $1000$ simulation runs of networks of size $N=600\thinspace000$ with mean degree $k=4$.}
\end{center}
\end{figure}

In order to illustrate this effect, in Fig.~\ref{fig:modules_at_s} we visualize the giant component at $S=0.1$ (close to total collapse) with interconnected nodes shown in black and all other nodes colored according to the module they belong to.
For a network with $m=4<m^*$, random node failure destroys the internal structure of the modules evenly, see Fig.~\ref{fig:modules_at_s}(a). 
In this random failure case, all the modules always appear in the giant component (i.e. there is always at least one node from each module in the giant component) as shown in Fig.~\ref{fig:modules_at_s}(e), and the size of modules is very narrowly distributed, see Fig.~\ref{fig:modules_at_s}(g). 
In contrast to random failure, when attacking the interconnected nodes (at $S=0.1$), see Fig.~\ref{fig:modules_at_s}(b), not all the modules remain in the giant component (for example, in Fig.~\ref{fig:modules_at_s}(b) there are only two of them). However, the modules that do remain, are almost intact, containing $14.6\%$ of their initial nodes, significantly more than in the random case. This point is demonstrated also in Supplementary Figs.~S5-S6 where we analyze the modules structure in the second largest cluster.

In contrast, for $m>m^*$, the interconnected nodes play an important role also in the internal structure of modules and therefore there is no need to remove all of them in order to break down the giant component.
Nevertheless, the attack still leaves them more complete than in the case of random removal, see Fig.~\ref{fig:modules_at_s}(c)-(d).
Furthermore, in the case of attack, usually \textit{all} modules appear in the giant component (see Fig.~\ref{fig:modules_at_s}(f)), and thus their relative size is smaller compared to the $m<m^*$ case (see Fig.~\ref{fig:modules_at_s}(g)). As $m$ increases, the difference between attack and random case becomes smaller, and as a result the percolation threshold converges to the one obtained for random failure, see Supplementary Fig.~S3.

For the case of $m<m^*$, the attack of interconnected nodes has a weak effect on the internal structure of the modules, and the removal of inter-module nodes results in an abrupt decrease in the size of the giant component, see Fig.~\ref{fig:inter_connected_failure}(b).
In addition, while for $m=100~>m^*$ (Fig.~\ref{fig:inter_connected_failure}(c)) we observe a regular second order percolation transition characterized by the continuous decrease of $S$ and the sharp peak in $S_{second}$, the case of $m<m^*$ demonstrates an abrupt, first order transitions. 
The reason is that the second largest cluster contains large connected subgraphs corresponding to modules who ``dropped'' from the giant component, see Fig.~\ref{fig:modules_at_s}(e). Therefore, with the emergence of the giant component, these modules become part of it, leading to a sudden drop in the size of the second largest cluster.

In Fig.~\ref{fig:inter_connected_failure}(d), we show the critical number of modules, $m^*$, as a function of $\alpha$ for networks with mean degree $k=4$. It is seen that $m^*$ is increasing with $\alpha$, and the percolation threshold at this point is ${p_c}^* \approx 0.3417$ independent of $\alpha$, meaning the transition takes place at a fixed inter-module average degree ${k_{\text{inter}}}^* = -\ln({p_c}^*) \approx 1.0738$. We show how the critical percolation threshold ${p_c}^*$ and the critical mean inter-degree ${k_{\text{inter}}}^*$ are changing with $k$ in Supplementary Fig.~S4.

In order to further demonstrate the transition in the $p_c$ behavior, in Fig.~\ref{fig:inter_connected_failure}(e) we show the percolation threshold as a function of $k_\text{inter}$ for networks with mean intra-degree $k_\text{intra}=2$ and number of modules $m=10$ fixed.
Here we see a similar transition in $p_c$ as before, but the critical point is now a function of the concentration of interconnected nodes.  
At a critical $k^*_\text{inter}~=~ k_\text{inter} \approx 0.693$, $p_c$ changes from Eq.~(\ref{eq:qc_1}) behavior to Eq.~(\ref{eq:qc_2}). 

Finally, in modular structures the interconnected nodes have high betweenness centrality (see Fig.~\ref{fig:bet}), and thus, our framework also provides an analytical tool of studying attacks on high betweenness centrality nodes, where only numerical simulations currently exist that suggest such an attack is one of the most harmful attack strategies~\cite{HolHKYH02,SchSMAHH11}.
Figure~\ref{fig:bet} compares the betweenness centrality of nodes with inter-module connections (called inter-nodes) and nodes with only intra-module connections (called intra-nodes) for networks of size $N=100\thinspace000$ with $m=10$ modules.
First, we show that the average betweenness centrality of interconnected nodes is significantly higher than for nodes without interconnections in networks with mean intra-degree $k_\text{intra} = 2$ and a varying number of interconnections, see Fig.~\ref{fig:bet}(a).
Then, for $k_\text{inter}=2$, we show that the betweenness centrality distribution of interconnected nodes has a broader tail, meaning that interconnected nodes are much more likely to have high betweenness centrality. In Supplementary Fig.~S7 we obtain similar results in networks with $k=4$ and different values of $\alpha$. Thus, our analytical results of attack on interconnected nodes can be regarded as a theory for attacking high betweenness nodes.

Our analytical and numerical investigation of the effect of modularity on network stability has important implications for real world networks, such as cognitive and neural brain networks. The modular architecture of neural structural and functional networks is considered a fundamental principle of the brain \cite{meunier2010modular}. This non-random modular architecture is crucial for the brain's functional demands of segregation and integration of information \cite{bullmore2012economy}. In fact, disrupted brain modular organization is related to neuropathology, such as schizophrenia \cite{alexander2010disrupted}, autism \cite{barttfeld2012state}, Alzheimer's \cite{van2012structure} and impulsivity \cite{davis2013impulsivity}. Nevertheless, research investigating any possible negative aspects of modular organization in brain networks is lacking. At the cognitive level (the level of information processing in the brain), network analysis is mainly focused on language and memory networks \cite{Baronchelli}. Yet, knowledge on modular effect and importance in cognitive network organization is limited. Recently, the semantic memory organization of persons with Asperger syndrome was compared to that of neurotypical controls using network analysis \cite{KenettAsperger}. This research found that the semantic memory network of persons with Asperger syndrome is more modular than that of neurotypical matched controls. The authors suggest that this ``hyper-modularity'' is related to the Asperger syndrome rigidity of thought, e.g. difficulty in comprehending high level aspects of language. Thus, modular organization can have a negative effect on real world networks by leading to rigidity of the network which might hinder proper network function.

Finally, our study offers an efficient immunization approach in modular networks, where epidemic spreading can be prevented at a lower cost by immunizing interconnected nodes.
For both regimes, below and above $m^*$, the percolation threshold obtained from attacking the interconnected nodes is higher than the case of random failure and therefore immunization of these nodes is more effective.
For the regime $m<m^*$, this can be done at a very low cost as the percolation threshold is very high.
Thus, in geographically distant social networks, it is worth vaccinating people that link between different communities such as businessmen traveling a lot between countries.

\section*{Methods}
We give a brief derivation of our analytical solution (Eqs.~(\ref{eq:qc_1})-(\ref{eq:s_i})). We extend Callaway \textit{et al.}'s approach~\cite{CalCNSW00} for studying the robustness of networks to intentional attacks, from single-module networks to modular networks in a similar manner that was done in~\cite{LeiLD09}. We define the generating functions for the degree and excess degree distributions of occupied nodes 
\begin{equation}
F_i(x_1, x_2, \ldots, x_m) = \sum\limits_{k_1, k_2, \ldots, k_m=0}^\infty {p^i}_{k_1, k_2, \ldots, k_m}  \ {r^i}_{k_1, k_2, \ldots, k_m} \ {x_1}^{k_1} {x_2}^{k_2} \ldots {x_m}^{k_m}
\label{eq:f_main}
\end{equation}
\begin{equation}
F_{ij}(x_1, x_2, \ldots, x_m) = \sum\limits_{k_1, k_2, \ldots, k_m=0}^\infty {q^{ij}}_{k_1, k_2, \ldots, k_m}  \ {r^i}_{k_1, k_2, \ldots, k_m} \ {x_1}^{k_1} {x_2}^{k_2} \ldots {x_m}^{k_m}
\label{eq:f_excess_main}
\end{equation}
where ${p^{i}}_{k_1, k_2, \ldots, k_m}$ is the probability that a node from module $i$ has degree $({k_1, k_2, \ldots, k_m})$, ${q^{ij}}_{k_1, k_2, \ldots, k_m}$ is the probability of following a randomly chosen $ij$-edge to a node with excess degree $(k_1, k_2, \ldots, k_m)$, and ${r^i}_{k_1, k_2, \ldots, k_m}$ is the occupation probability of a node with degree $(k_1, k_2, \ldots, k_m)$ defined in (\ref{eq:occ_prob}).
By substituting (\ref{eq:occ_prob}) into (\ref{eq:f_main})-(\ref{eq:f_excess_main}), in the case of modular ER networks with average intra- and inter-degree $k_{\text{intra}}$, $k_{\text{inter}}$ respectively, we obtain 
\begin{equation}
F_i(x) = e^{k_{\text{intra}}(x_i - 1) - k_{\text{inter}}} (1-q) + q G_i(x)
\label{eq:final_fi}
\end{equation}
\begin{equation}
F_{ij}(x) = 
\begin{cases}
    \frac{\partial F_i}{\partial x_j}(x) \frac{m-1}{k_{\text{inter}}}  = q G_i(x),& \text{if } i \neq j\\
    \frac{\partial F_i}{\partial x_i}(x) \frac{1}{k_{\text{intra}}}  = F_i(x),& \text{otherwise}
\end{cases}
\label{eq:final_fij}
\end{equation}
where $x=(x_1, x_2, \ldots, x_m)$ and $G_i(x)$ is the generating function of the degree distribution, see Supplementary Information.
We define the generating function for the distribution of the number of occupied nodes in the component reachable by following a randomly chosen $ij$-edge to a $i$-node and then following its additional outgoing links
\begin{equation}
J_{i j}(x) = 1 - F_{i j}(1) + x_i F_{i j}[J_{1i}, J_{2i}, \ldots, J_{mi}].
\label{eq:hexcessmain}
\end{equation}
And similarly, the distribution of the number of nodes reachable from a randomly chosen $i$-node (rather than $ij$-edge) is generated by
\begin{equation}
J_{i}(x) = 1 - F_{i}(1) + x_i F_{i}[J_{1i}, J_{2i}, \ldots, J_{mi}].
\label{eq:hmain}
\end{equation}
Then, the average number of occupant $j$-nodes in the component of a randomly chosen $i$-node, is given by
\begin{align}
\label{eq:sijinter}
{\langle s_i \rangle}_j &=  \frac{\partial J_{i}}{\partial x_j}(x) |_{x=1}   \nonumber \\
&= {\delta}_{i j} F_{i}(1) +  \frac{\partial F_{i}}{\partial x_j}(1) \frac{\partial J_{i i}}{\partial x_j}(1) +  \sum_{\substack{l=1 \\ l \neq i}}^m \frac{\partial F_{i}}{\partial x_l}(1) \frac{\partial J_{li}}{\partial x_j}(1)  \nonumber \\
&= {\delta}_{ij} F_{i}(1) + k_{\text{intra}} F_i(1) \frac{\partial J_{ii}}{\partial x_j}(1) + q \frac{k_{\text{inter}}}{m-1} \sum_{\substack{l=1 \\ l \neq i}}^m \frac{\partial J_{li}}{\partial x_j}(1).
\end{align}
Solving the system~(\ref{eq:sijinter}), see full details in the Supplementary Information, we obtain the critical occupation probability of interconnected nodes in which the average component size diverges, given in Eqs.~(\ref{eq:qc_1})-(\ref{eq:qc_2}).
Finally, once the giant component emerges ($p > p_c$), the fraction of $i$-nodes belonging to the giant component, $S_i$, is given by
\begin{equation}
S_i =  1 - J_i(1) = F_i(1) - F_i(u_{1i}, u_{2i}, \ldots, u_{mi})
\label{eq:Siattack}
\end{equation}
where $u_{ji} = 1-S_j$, yielding Eq.~(\ref{eq:s_i}).

%\clearpage 

\bigskip
\bigskip

\noindent{\it Acknowledgments}

\noindent SH and DYK thank DTRA, ONR, BSF, the LINC (No. 289447) and the Multiplex (No. 317532) EU projects, the DFG, and the Israel Science Foundation for support. SS is supported by a scholarship from the Scottish Informatics and Computer Science Alliance.

\bigskip

\noindent{\it Author contributions}

\noindent SS, DYK, YNK, SD, MF, and SH performed the research and wrote
the paper.

\bigskip

%\noindent{\it Additional information}

%\noindent Supplementary information is available in the online version of the paper.
%Reprints and permissions information is available online at www.nature.com/reprints. Correspondence and requests for materials should be addressed to S.S.

\bigskip

\noindent{\it Competing financial interests}

\noindent The authors declare no competing financial interests. 

\bigskip

%\bibliography{bib}


\clearpage 

\begin{figure}[h]
\begin{center}
\subfigure{\includegraphics[scale=0.45]{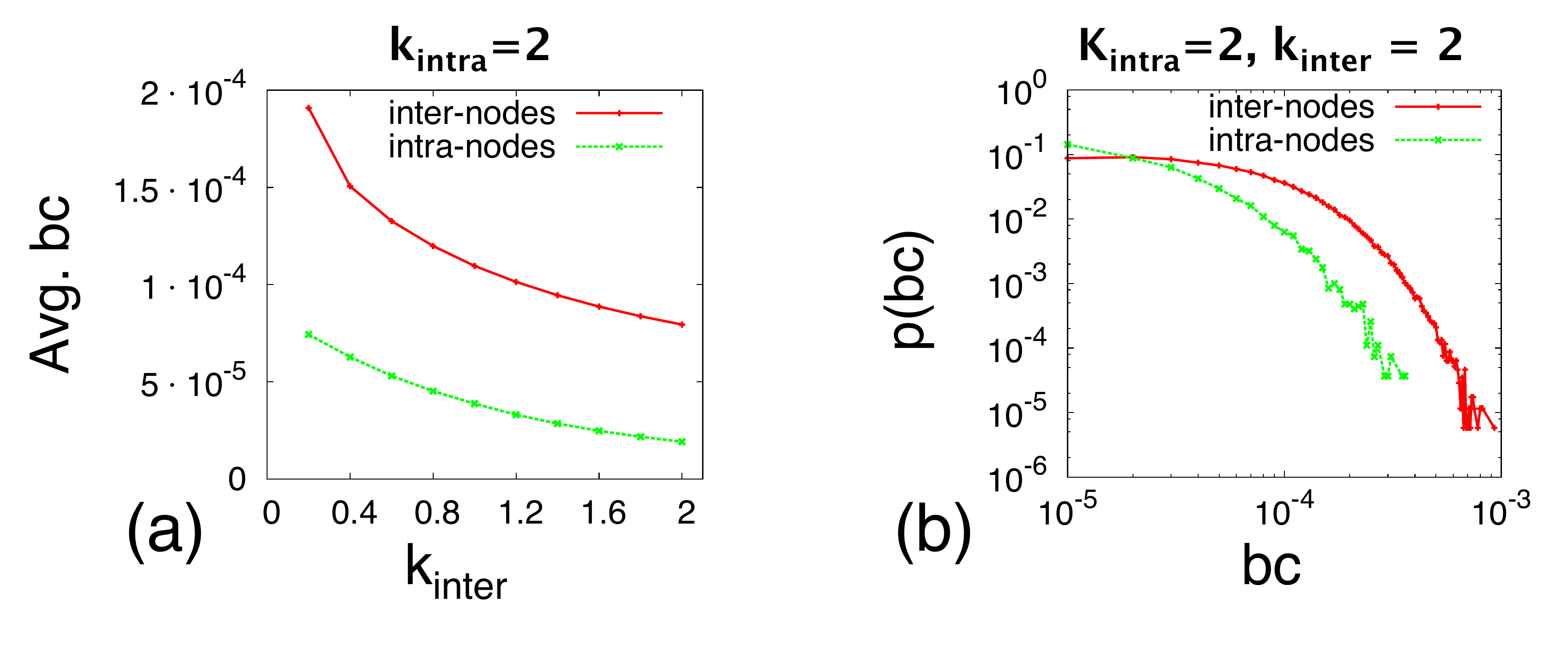}}
\caption{\label{fig:bet} {\bf Betweenness centrality of interconnected nodes.} Betweenness centrality of inter-nodes (nodes that have at least one interconnection) and intra-nodes (nodes with only intraconnections) in networks of size $N=100\thinspace000$ with $m=10$ modules. (a) Mean betweenness centrality as a function of $k_\text{inter}$ in networks with $k_\text{intra}=2$. (b) Distribution of betweenness centrality in networks with $k_\text{intra} = k_\text{inter} = 2$.}
\end{center}
\end{figure}

%%%%%%%%%%%%%%% SI %%%%%%%%%%%%%%%%%%%%%

\clearpage

\noindent
{\LARGE Resilience of modular complex networks:\\ Supplementary Information} \\[0.2in]
%\noindent
{\large Saray Shai, Dror Y. Kenett, Yoed N. Kenett, Miriam Faust, Simon Dobson, and Shlomo Havlin}

\makeatletter \renewcommand{\thefigure}{S\@arabic\c@figure} \renewcommand{\thetable}{S\@arabic\c@table} \makeatother

\section{Introduction}
We present supplementary material on our paper: ``Resilience of modular complex networks''. First, in section~\ref{sec:analytical} we describe in more details the derivation of the analytical solution presented in the main text. Then, in section~\ref{sec:model}, we examine the properties of our model for generating random modular networks. In section~\ref{sec:perc} we further discuss the two percolation regimes found in the analytical framework and their implications. In section~\ref{sec:second} we examine the internal structure of modules in the second largest cluster, and finally in section~\ref{sec:bc} we discuss the betweenness centrality of interconnected nodes in modular structures.

\section{Analytical solution}
\label{sec:analytical}
\subsection{Random failure}
First, we give full derivation of Eq.~(5) in the main text. The generating function for the degree and excess degree distribution (see~\cite{SUPLeiLD09}) of modular ER networks with average intra- and inter-degree $k_{\text{intra}}$, $k_{\text{inter}}$ respectively is given by 
\begin{equation}
G_i(x) =  e^{k_{\text{intra}} (x_i - 1)} e^{  \frac{k_{\text{inter}}}{m-1} \sum\limits_{j \neq i} (x_j - 1)}
\tag{S1}\label{eq:g_main_exp}
\end{equation}
\begin{equation}
G_{ij}(x) = 
\begin{cases}
    \frac{\frac{\partial G_i}{\partial x_j}(x)}{\frac{\partial G_i}{\partial x_j}(1)}  =  \frac{k_{\text{inter}}}{m-1} G_i(x) \frac{m-1}{ k_{\text{inter}}} = G_i(x),& \text{if } i \neq j\\\\[-16pt]
    \frac{\frac{\partial G_i}{\partial x_j}(x)}{\frac{\partial G_i}{\partial x_j}(1)}  = k_{\text{intra}} G_i(x) \frac{1}{k_{\text{intra}}} = G_i(x), & \text{otherwise}
\end{cases}
\tag{S2}\label{eq:g_excess_exp}
\end{equation}
where $x=(x_1, x_2, \ldots, x_m)$.
Then from~\cite{SUPLeiLD09}, the average number of $j$-nodes in the component of a randomly chosen $i$-node is given by
\begin{equation}
{\langle s_i \rangle}_j = {\delta}_{ij} + k_{\text{intra}} \frac{\partial H_i}{\partial x_j}(1) + \frac{k_{\text{inter}}}{m-1} \sum_{\substack{l=1 \\ l \neq i}}^m \frac{\partial H_l}{\partial x_j}(1)
\tag{S3}\label{eq:s_ij}
\end{equation}
where ${\delta}_{ij}$ denotes the Kronecker delta and 
\begin{equation}
H_{ij}(x) = x_i G_{ij}[H_{1i}, H_{2i}, \ldots, H_{mi}] \ \ \ ; \ \ \ H_{i}(x) = x_i G_{i}[H_{1i}, H_{2i}, \ldots, H_{mi}]
\tag{S4}\label{eq:h_01}
\end{equation}
For example, using the notation $h_i = \frac{\partial H_i}{\partial x_1}(1)$, the system obtained for ${\langle s_1 \rangle}_1$ is:
\begin{align}
h_1 &= 1 + k_\text{intra} h_1 + \frac{k_\text{inter}}{m-1} (h_2 + h_3 + \ldots + h_m) \nonumber \\
h_2 &= k_\text{intra} h_2 + \frac{k_\text{inter}}{m-1} (h_1 + h_3 + \ldots + h_m) \nonumber \\
\vdots \nonumber \\
h_m &= k_\text{intra} h_m + \frac{k_\text{inter}}{m-1} (h_1 + h_2 + \ldots + h_{m-1}) 
\tag{S5}\label{eq:h1er}
\end{align}
Summing equations for $h_2, \ldots h_m$, we obtain:
\begin{align}
& h_2 + h_3 + ... + \dots h_m = k_\text{inter} h_1  + (k_\text{intra} + \frac{m-2}{m-1} {k_\text{inter}}(h_2 + h_3 + \ldots h_m) \nonumber \\ 
&\Rightarrow  h_2 + h_3 + \ldots h_m = \frac{k_\text{inter} h_1}{1 - k_\text{intra} - \frac{m-2}{m-1} {k_\text{inter}}}
\tag{S6}\label{eq:sum}
\end{align}
Substituting into~(\ref{eq:h1er}), we obtain:
\begin{align}
& h_1 = 1 + k_\text{intra} h_1 + \frac{k_\text{inter}}{m-1} (h_2 + h_3 + \ldots + h_m) \nonumber \\
= & 1 + k_\text{intra} h_1 + \frac{k_\text{inter}}{m-1} (\frac{k_\text{inter} h_1}{1 - k_\text{intra} - \frac{m-2}{m-1} {k_\text{inter}}}) \nonumber \\ 
&\Rightarrow h_1(1 - k_\text{intra} - \frac{ \frac{k_\text{inter}^2}{m-1}}{1 - k_\text{intra} - \frac{m-2}{m-1} {k_\text{inter}}}) = 1 \nonumber  \\[3ex]
&\Rightarrow h_1 = \frac{1 - k_\text{intra} - \frac{m-2}{m-1} {k_\text{inter}}}{(1 - k_{\text{intra}})(1 - k_{\text{intra}} - \frac{m-2}{m-1} {k_\text{inter}}) - \frac{{k_{\text{inter}}}^2}{m-1}}.
\tag{S7}\label{eq:h1_final}
\end{align}
$h_1$ diverges when $(1 - k_{\text{intra}})(1 - k_{\text{intra}} - \frac{(m-2)k_{\text{inter}}}{m-1}) - \frac{{k_{\text{inter}}}^2}{m-1} = 0$. This is also where all $h_i$ diverges, yielding Eq.~(5) in the main text.

\subsection{Attack on interconnected nodes}

In the case of attack on interconnected nodes, the average number of $j$-nodes in the component of a randomly chosen $i$-node is given by
\begin{equation}
\tag{S8}\label{eq:supsijinter}
{\langle s_i \rangle}_j = {\delta}_{ij} F_{i}(1) + k_{\text{intra}} F_i(1) \frac{\partial J_{ii}}{\partial x_j}(1) + q \frac{k_{\text{inter}}}{m-1} \sum_{\substack{l=1 \\ l \neq i}}^m \frac{\partial J_{li}}{\partial x_j}(1).
\end{equation}
where 
\begin{align}
F_i(x) &= e^{k_{\text{intra}}(x_i - 1) - k_{\text{inter}}} (1-q) + q G_i(x) \nonumber \\ 
F_{ij}(x) &=   \begin{cases}
   \frac{\partial F_i}{\partial x_j}(x) \frac{m-1}{k_{\text{inter}}}  = q G_i(x),& \text{if } i \neq j\\
   \frac{\partial F_i}{\partial x_i}(x) \frac{1}{k_{\text{intra}}}  = F_i(x),& \text{otherwise}
\end{cases} \nonumber \\ 
J_{i j}(x) &= 1 - F_{i j}(1) + x_i F_{i j}[J_{1i}, J_{2i}, \ldots, J_{mi}] \nonumber \\ 
J_{i}(x) &= 1 - F_{i}(1) + x_i F_{i}[J_{1i}, J_{2i}, \ldots, J_{mi}] 
\tag{S9}\label{eq:fj}
\end{align}

For example, using the notation $j_{ij} = \frac{\partial J_{ij}}{\partial x_1}(1)$, the system obtained for ${\langle s_1 \rangle}_1$ is:
\begin{align}
j_{11} &= F_1(1) + k_\text{intra} F_1(1) j_{11} + q \frac{k_\text{inter}}{m-1} ( j_{21} + j_{31} + \ldots + j_{m1}) \nonumber \\
j_{12} &=  q + q \ k_\text{intra} \ j_{11} + q \frac{k_\text{inter}}{m-1} ( j_{21} + j_{31} + \ldots + j_{m1}) \nonumber \\
\vdots \nonumber \\
j_{1m} &= q + q \ k_\text{intra} \ j_{11} + q \frac{k_\text{inter}}{m-1} ( j_{21} + j_{31} + \ldots + j_{m1}) \nonumber \\
j_{21} &= q \ k_\text{intra} \ j_{22} +  q \frac{k_\text{inter}}{m-1} ( j_{12} + j_{32} + \ldots + j_{m2}) \nonumber \\
j_{22} &= k_\text{intra} F_2(1) j_{22} +  q \frac{k_\text{inter}}{m-1} ( j_{12} + j_{32} + \ldots + j_{m2}) \nonumber \\
\vdots \nonumber \\
j_{2m} &= q \ k_\text{intra} \ j_{22} +  q \frac{k_\text{inter}}{m-1} ( j_{12} + j_{32} + \ldots + j_{m2}) \nonumber \\
\vdots \nonumber \\
j_{m1} &= q \ k_\text{intra} \ j_{mm} +  q \frac{k_\text{inter}}{m-1} ( j_{1m} + j_{3m} + \ldots + j_{m-1 m}) \nonumber \\
j_{m2} &= q \ k_\text{intra} \ j_{mm} +  q \frac{k_\text{inter}}{m-1} ( j_{1m} + j_{3m} + \ldots + j_{m-1 m}) \nonumber \\
\vdots \nonumber \\
j_{mm} &= k_\text{intra} F_m(1) j_{mm} +  q \frac{k_\text{inter}}{m-1} ( j_{1m} + j_{3m} + \ldots + j_{m-1 m}) \nonumber \\
\tag{S10}\label{eq:j11er}
\end{align}
Since $F_1(1) = F_i(1) = e^{-k_\text{inter}} (1-q) + q $ for all $i$, we can sum all equations for $j_{ii}$ obtaining:
\begin{align}
j_{11} + j_{22} +  \dots + j_{mm} &= F_1(1) + k_\text{intra} F_1(1) (j_{11} + j_{22} +  \dots + j_{mm}) + \nonumber \\
&+ q \frac{k_\text{inter}}{m-1} (j_{12} + j_{13} + \ldots + j_{1m} + \ldots + j_{m1} + j_{m2} + \ldots + j_{m m-1}) 
\tag{S11}\label{eq:sumjii}
\end{align}

Summing for all equations $j_{il}$ for $i \neq l$ we obtain
\begin{align}
& j_{12} + j_{13} + \ldots + j_{1m} + \ldots + j_{m1} + j_{m2} + \ldots + j_{m m-1} = (m-1)q + (m-1) \ q \ k_\text{intra} (j_{11} +  \dots + j_{mm}) + \nonumber \\
&+ q \ k_\text{inter}  (j_{12} + j_{13} + \ldots + j_{1m} + \ldots + j_{m1} + j_{m2} + \ldots + j_{m m-1}) \nonumber  \\[2ex]
& \Rightarrow j_{12} + j_{13} + \ldots + j_{1m} + \ldots + j_{m1} + j_{m2} + \ldots + j_{m m-1} = \frac{(m-1)q + (m-1) \ q \ k_\text{intra} (j_{11} + \dots + j_{mm})}{1 - q \ k_\text{inter}}
\tag{S12}\label{eq:sumjij}
\end{align}
And by substituting (\ref{eq:sumjij}) into (\ref{eq:sumjii}), we obtain
\begin{align}
j_{11} +  \dots + j_{mm} &=  F_1(1) + k_\text{intra} F_1(1) + \frac{q^2 \ k_\text{inter}}{1 - q \ k_\text{inter}} + \frac{q^2 \ k_\text{intra} \ k_\text{inter} (j_{11} +  \dots + j_{mm})}{1 - q \ k_\text{inter}}  \nonumber \\[2ex]
& \Rightarrow j_{11} +  \dots + j_{mm} = \frac{F_1(1) (1 - q \ k_\text{inter}) + q^2 \ k_\text{inter}}{(1 - k_\text{intra} F_1(1))(1 - q \ k_\text{inter}) - q^2 \ k_\text{intra} \ k_\text{inter}}
\tag{S13}\label{eq:sumjiifinal}
\end{align}
leading to the critical occupation probability of interconnected nodes (in which the average component size diverges) given in Eqs.~(8)-(9) in the main text.

\clearpage 

\section{Model for generating random modular networks}
\label{sec:model}

\begin{figure}[h]
\begin{center}
\subfigure{\label{fig:vis}\includegraphics[scale=0.42]{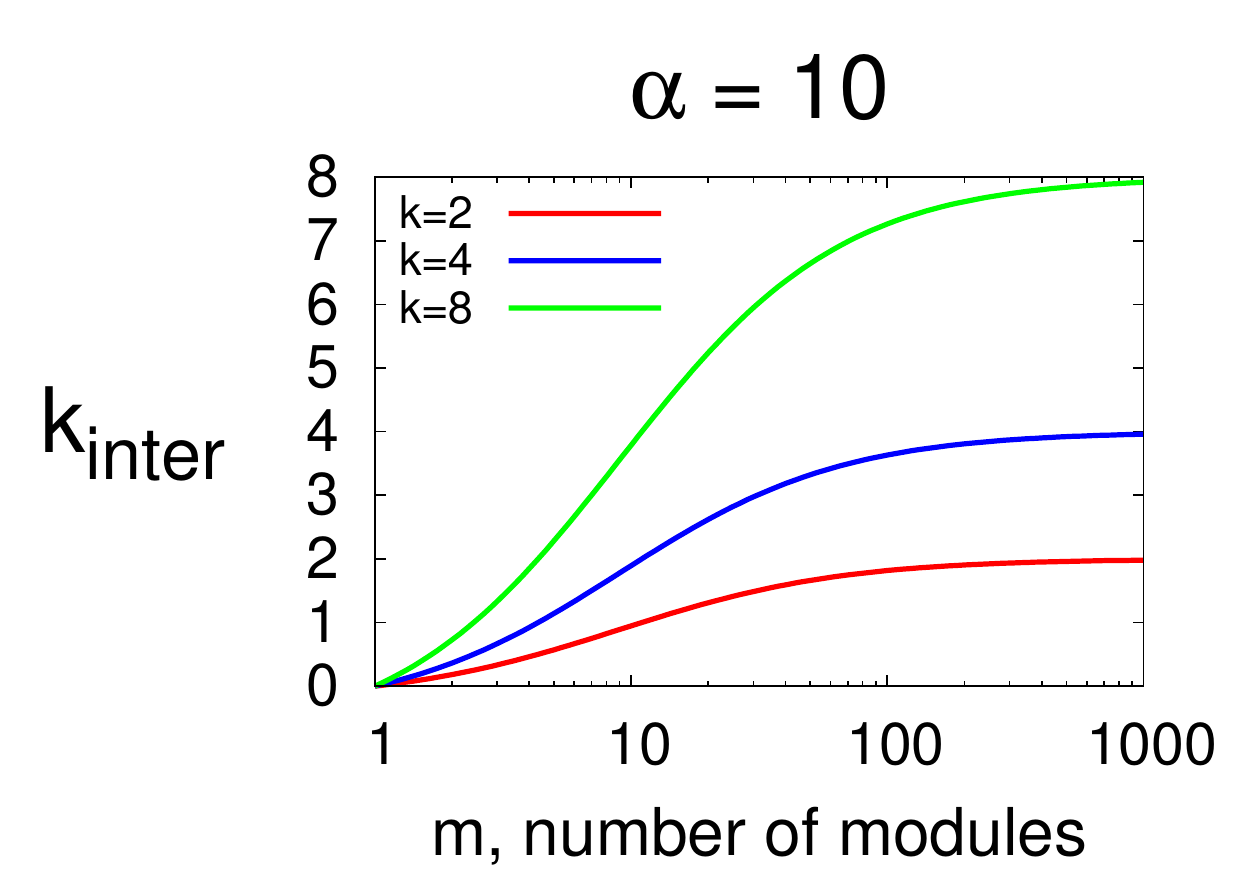}}
\subfigure{\label{fig:vis}\includegraphics[scale=0.42]{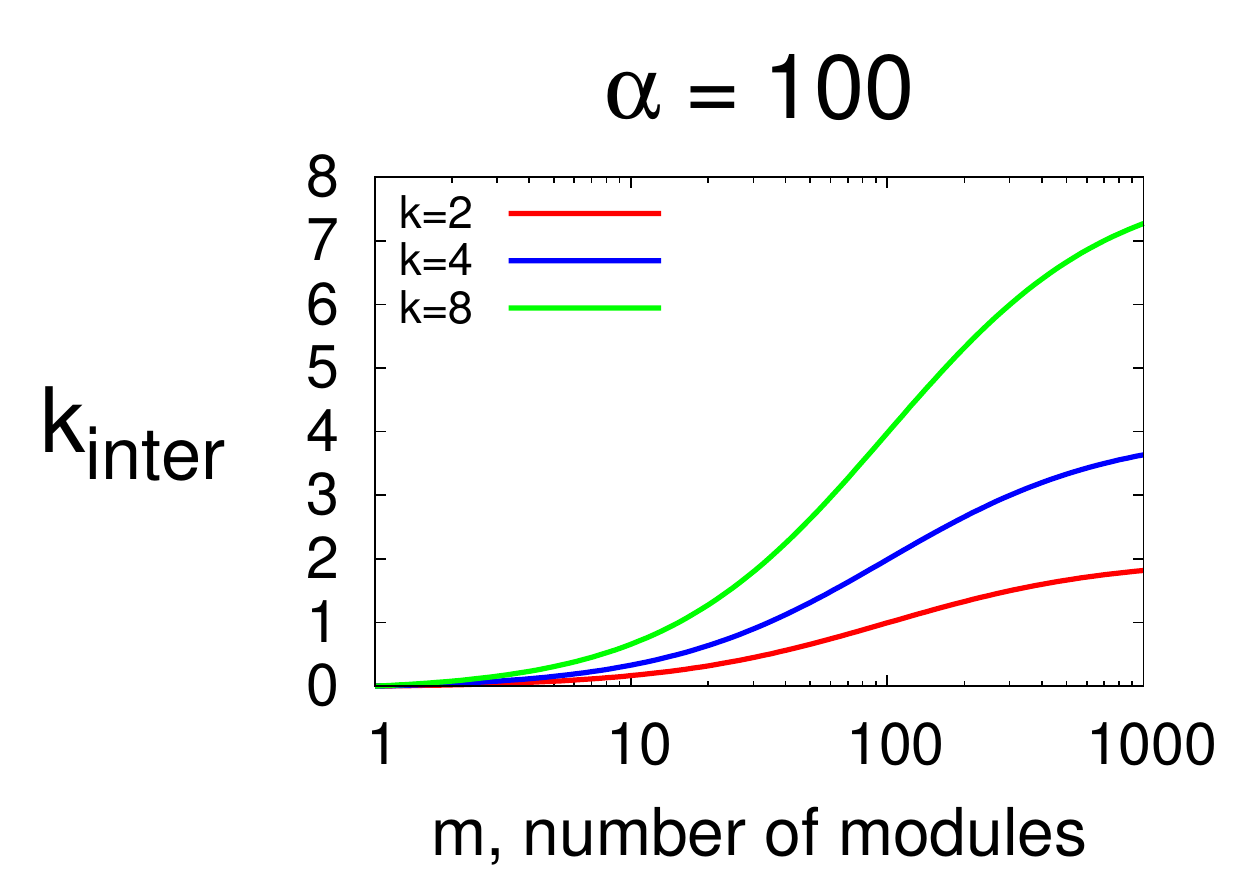}}
\subfigure{\label{fig:vis}\includegraphics[scale=0.42]{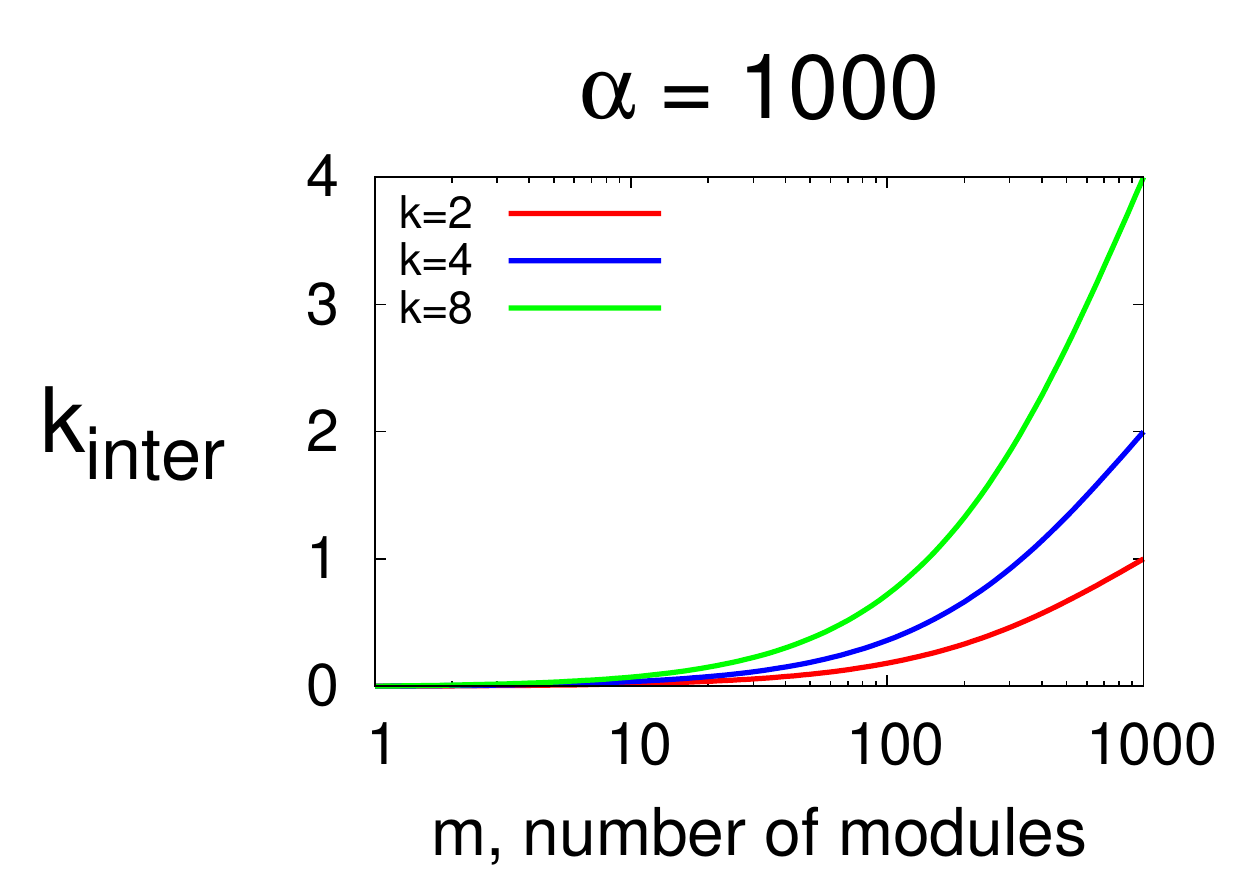}}
\caption{\label{fig:k_inter} {\bf The effect of \boldmath{$\alpha$} on the convergence of \boldmath{$k_\text{inter}$} to the mean degree.} According to the model described in the main text, $k_\text{inter}$ is increasing with $m$ taking into consideration that systems comprised of more modules have more inter-links accordingly. At the limit of large $m$, $k_\text{inter}$ is approaching the mean degree in a rate determined by $\alpha$.}
\end{center}
\end{figure}

\clearpage

\section{Two percolation regimes}
\label{sec:perc}

\begin{figure}[h]
\begin{center}
\subfigure{\includegraphics[scale=0.55]{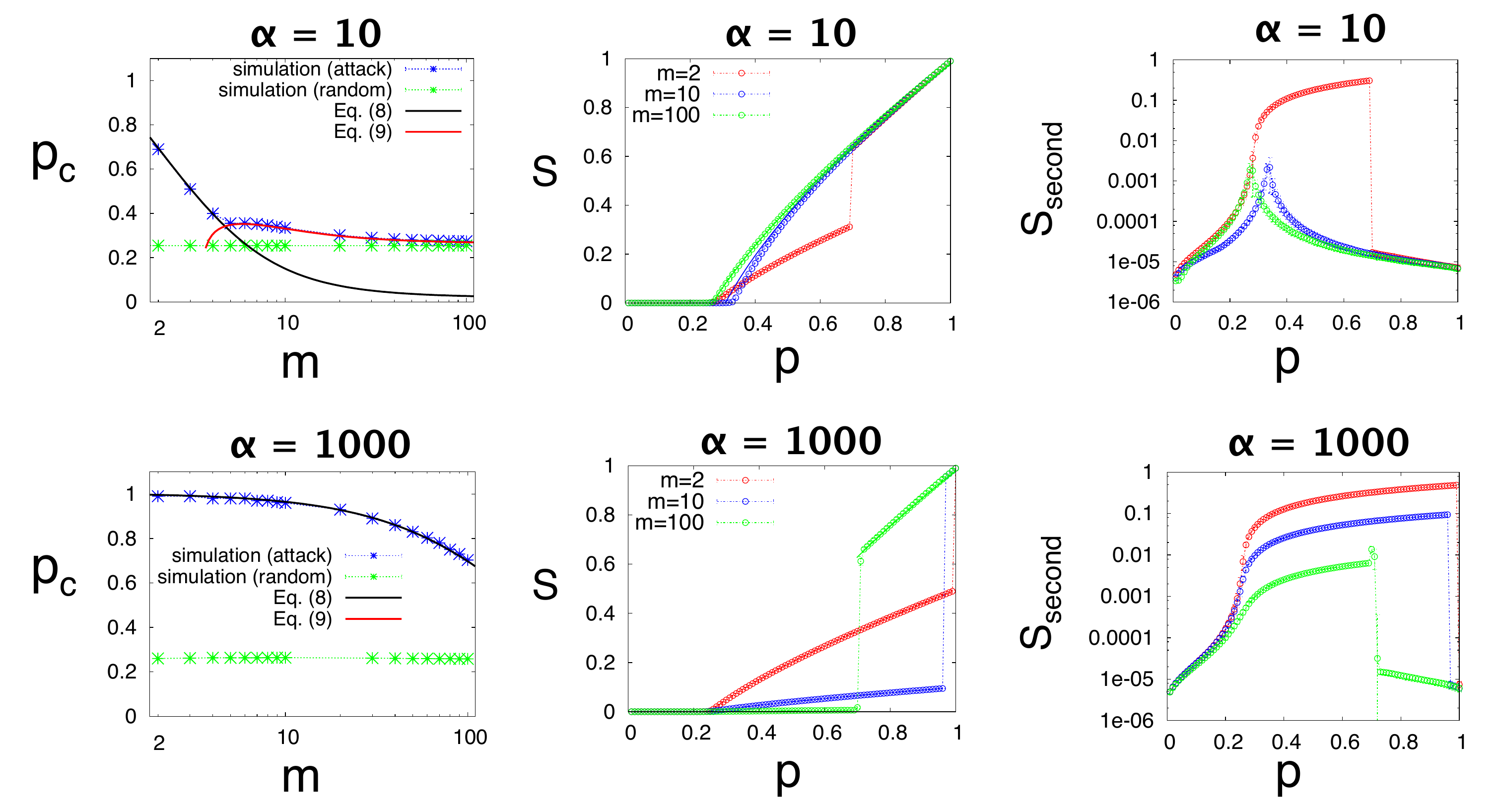}}
\caption{\label{fig:giant_component} {\bf Two percolation regimes when attacking interconnected nodes.} Here we show the two percolation regimes discussed in the main text for $\alpha=10$ and $\alpha=1000$. Results shown are for networks with fixed mean degree $k=4$. As discussed in the main text, the regime of $m<m^*$, corresponds to $q_c=0$, is characterized by an abrupt first order transition, while for $m>m^*$ we observe a regular second order percolation transition characterized by the continuous decrease of $S$ and the sharp peak in $S_{second}$.}
\end{center}
\end{figure}

\begin{figure}[h]
\begin{center}
\subfigure{\includegraphics[scale=0.42]{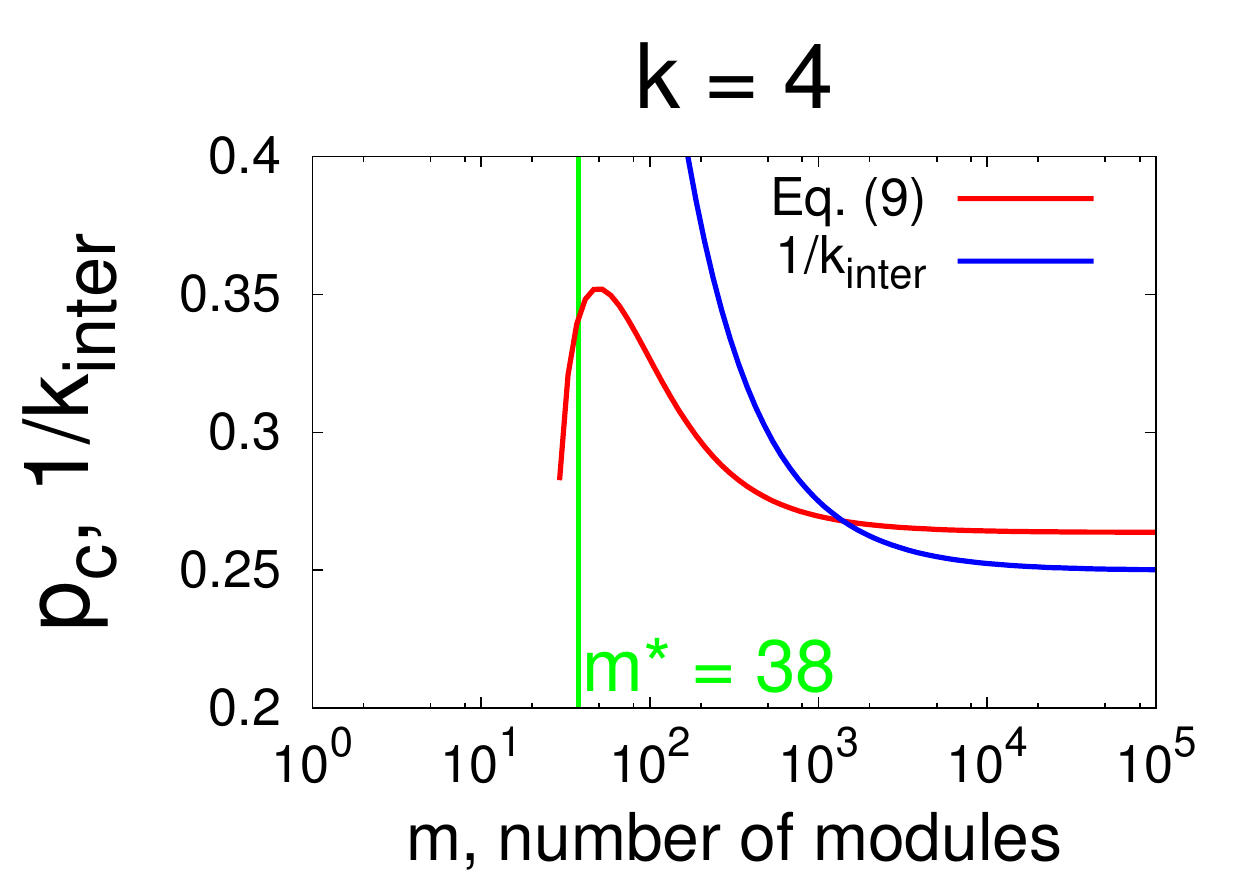}}
\subfigure{\includegraphics[scale=0.42]{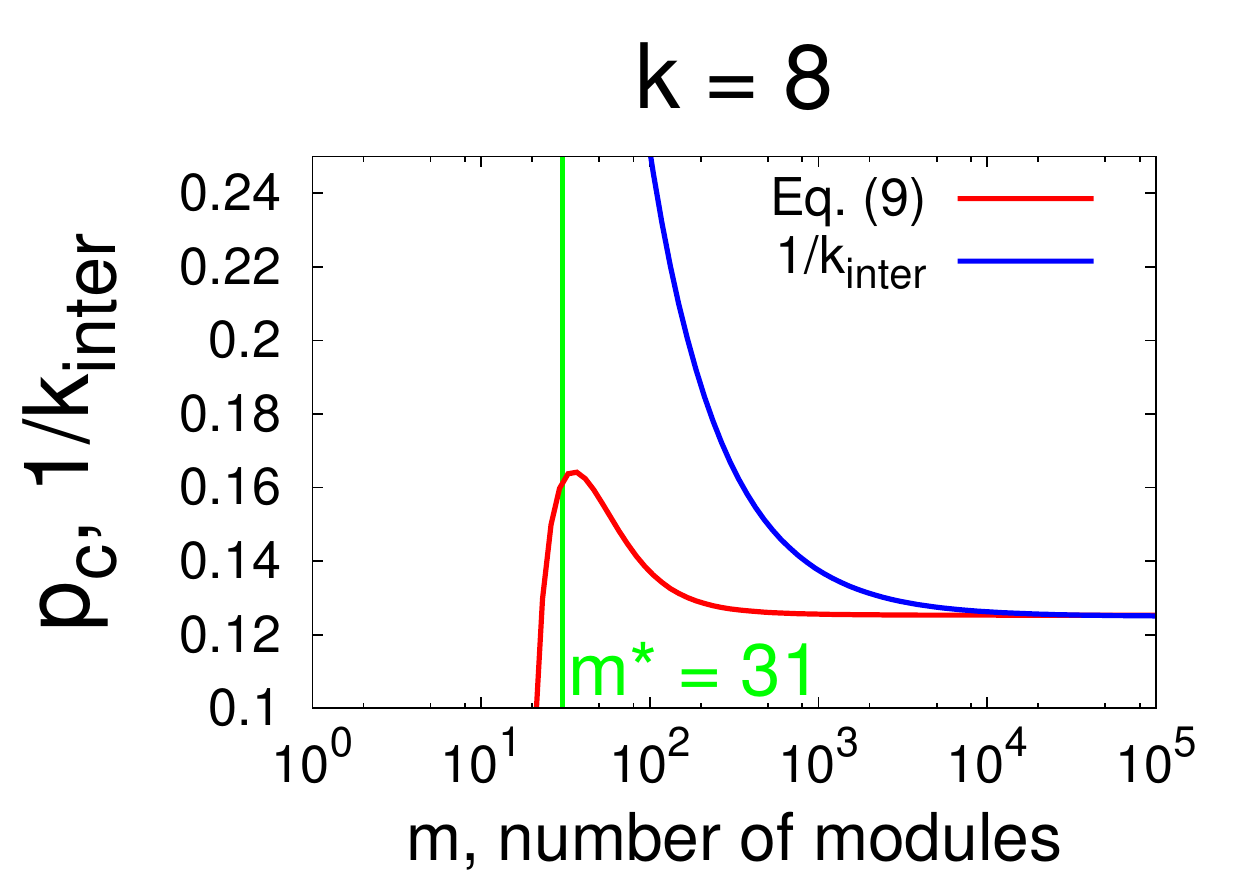}}
\subfigure{\includegraphics[scale=0.42]{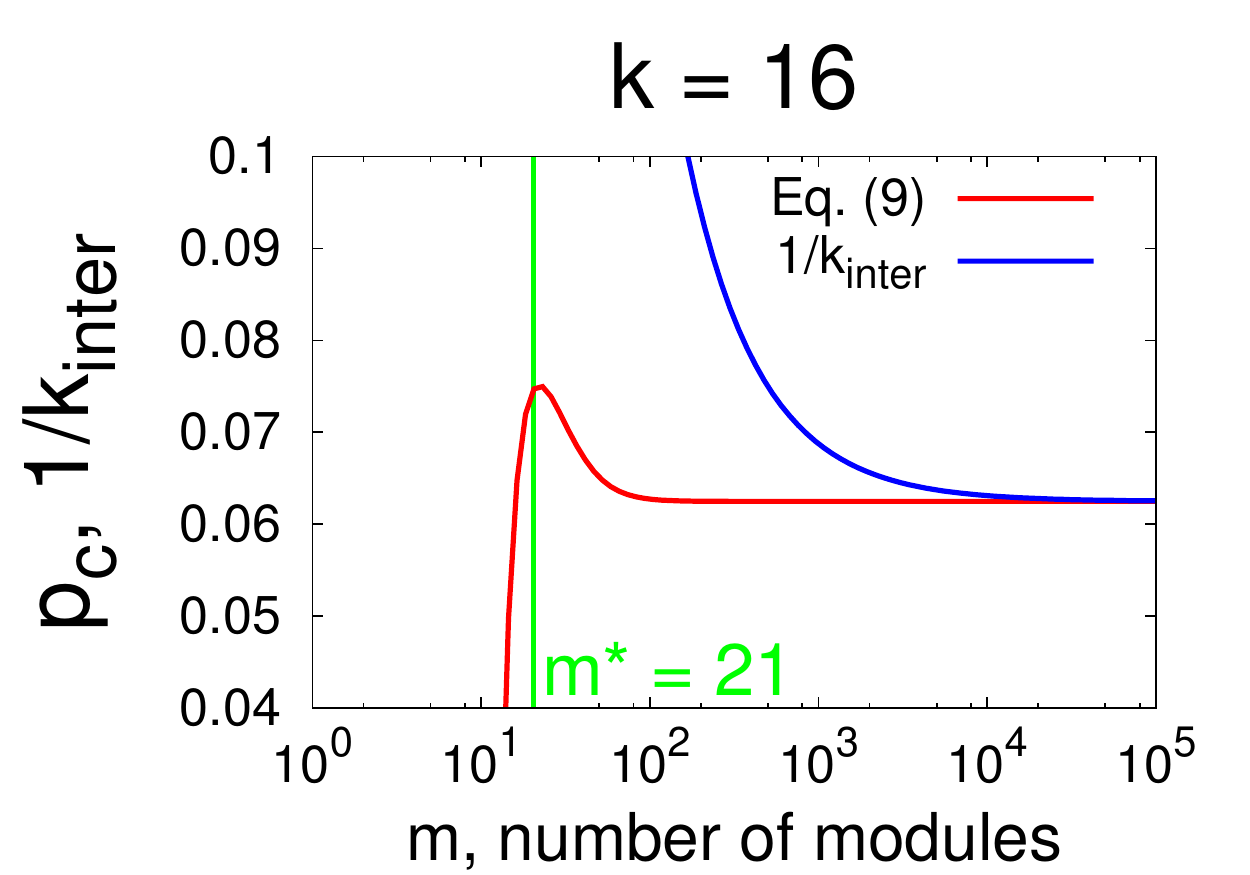}}
\caption{\label{fig:pc_converge} {\bf Convergence of \boldmath{$p_c$} to \boldmath{$1 / k_\text{inter}$}.} Analytical results for $\alpha=100$. Red lines represent $p_c$ obtained from Eq.~(9) in the main text, blue lines shows $1/k_\text{inter}$ and green lines shows $m^*$. Since unlike random node removal, nodes with no links are never removed in the attack, $p_c$ is converging to $1/k + e^{-k}$. Here we show that indeed for higher degrees than shown in the main text, our model is converging to the percolation threshold of random removal as the number of modules increases.} 
\end{center}
\end{figure}

\begin{figure}[h]
\begin{center}
\subfigure{\label{fig:vis}\includegraphics[scale=0.6]{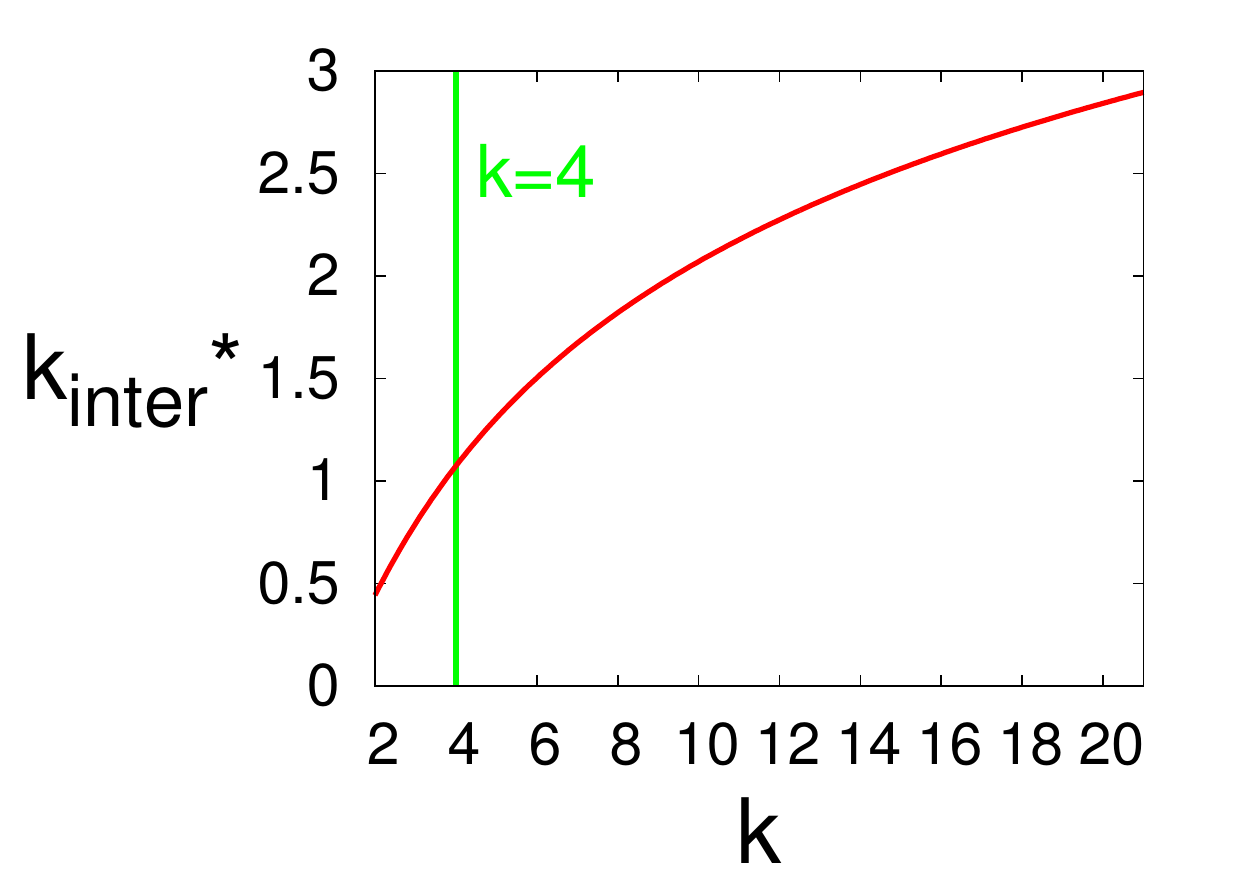}}
\subfigure{\label{fig:vis}\includegraphics[scale=0.6]{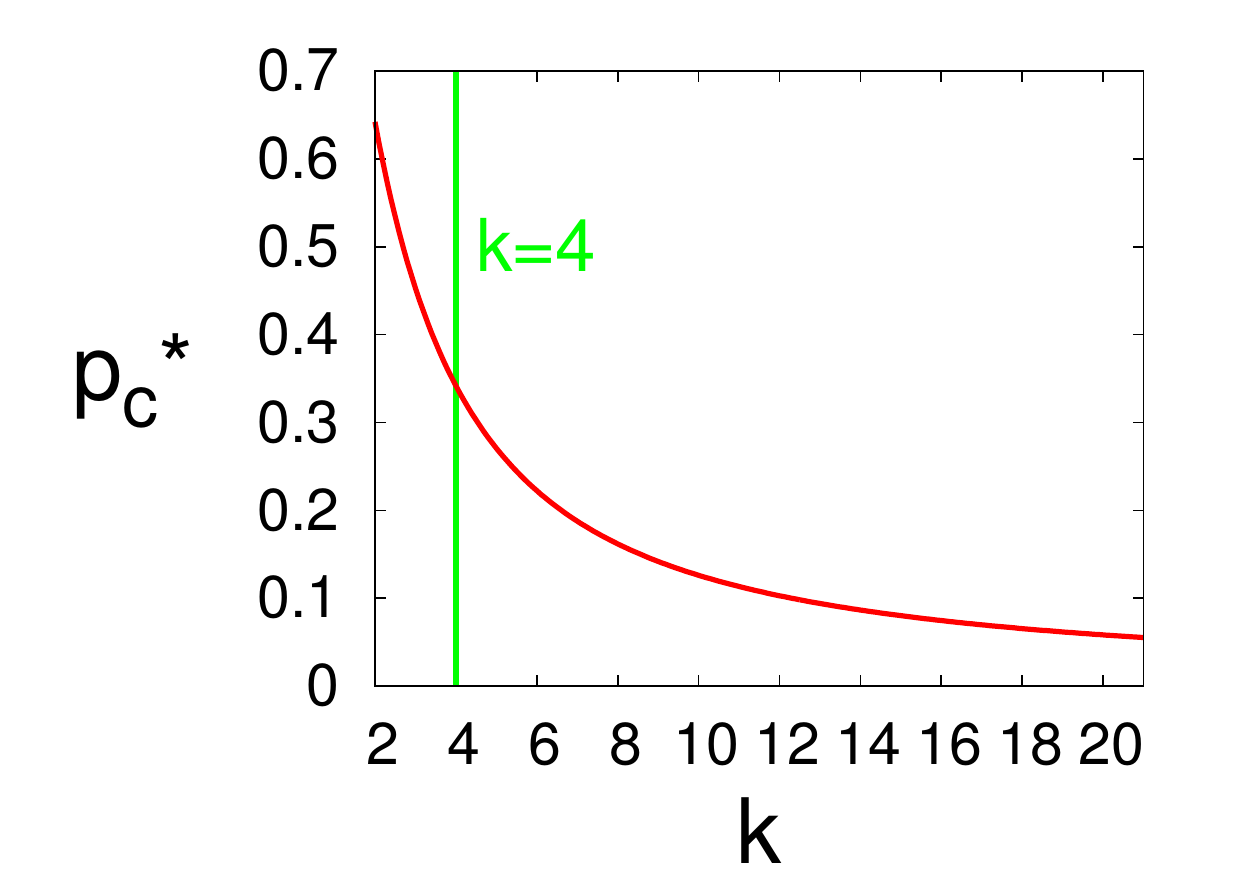}}
\caption{\label{fig:star_vs_k} {\bf Changing of \boldmath{$p_c^*$} and \boldmath{$k_\text{inter}^*$} as the mean degree increases.} The percolation threshold $p_c^*$ and the mean inter-degree $k_\text{inter}^*$ at the transition point between the two regimes (where the two solutions for $p_c$ cross) as a function of $k$. In the main text we show results for $k=4$ where $p_c^* \approx 0.3417$ and $k_\text{inter}^* \approx 1.0738$ (vertical lines).}
\end{center}
\end{figure}

\clearpage

\section{Modules structure in the second largest cluster}
\label{sec:second}

\begin{figure}[h]
\begin{center}
\subfigure{\includegraphics[scale=0.42]{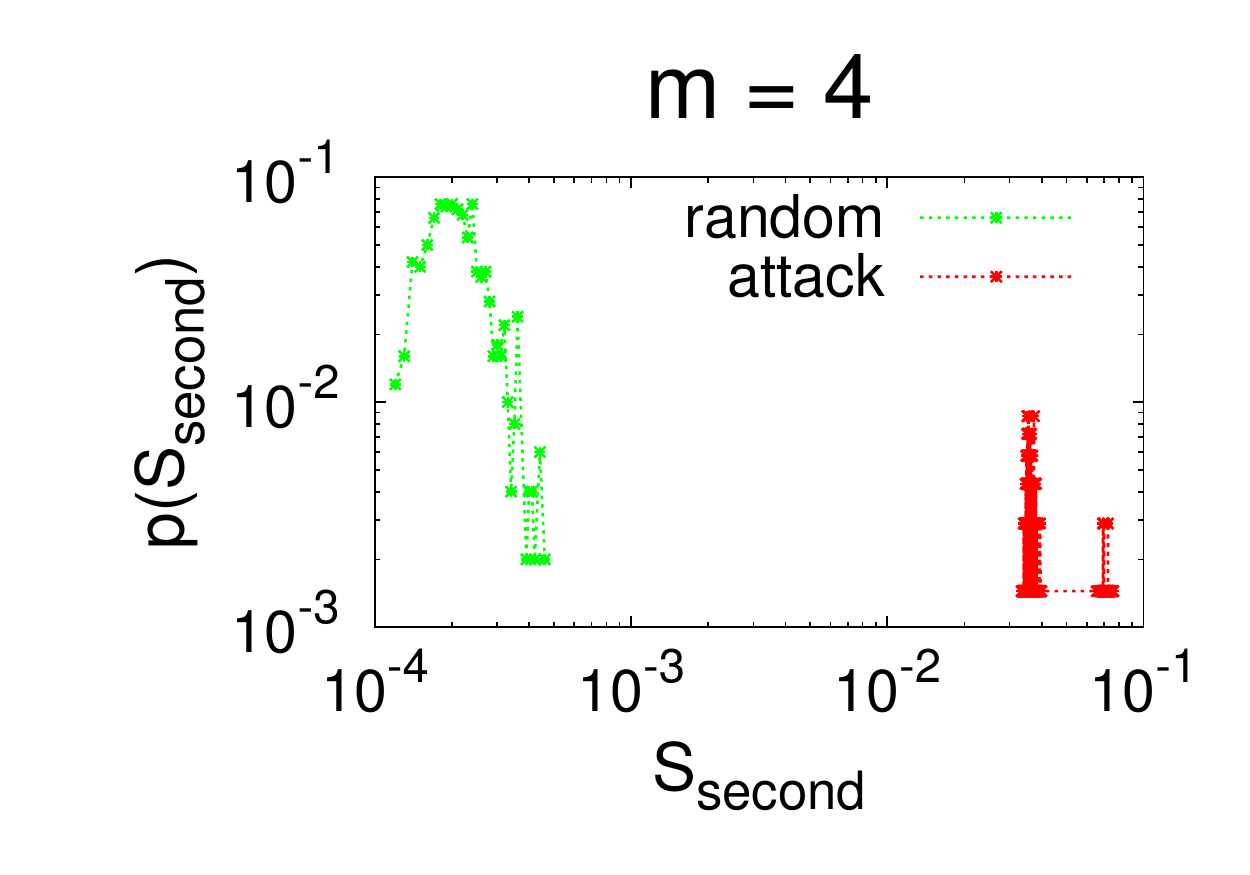}}
\subfigure{\includegraphics[scale=0.42]{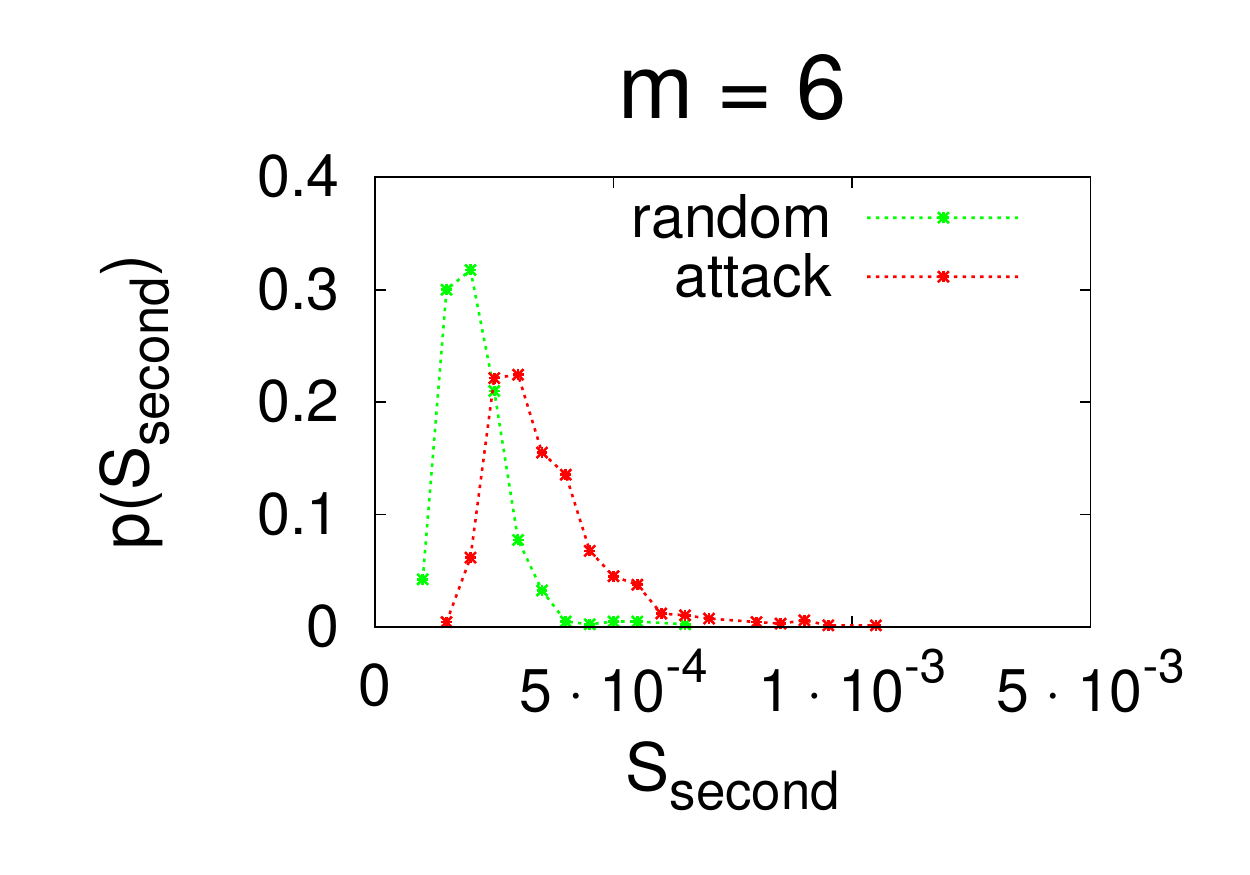}}
\subfigure{\includegraphics[scale=0.42]{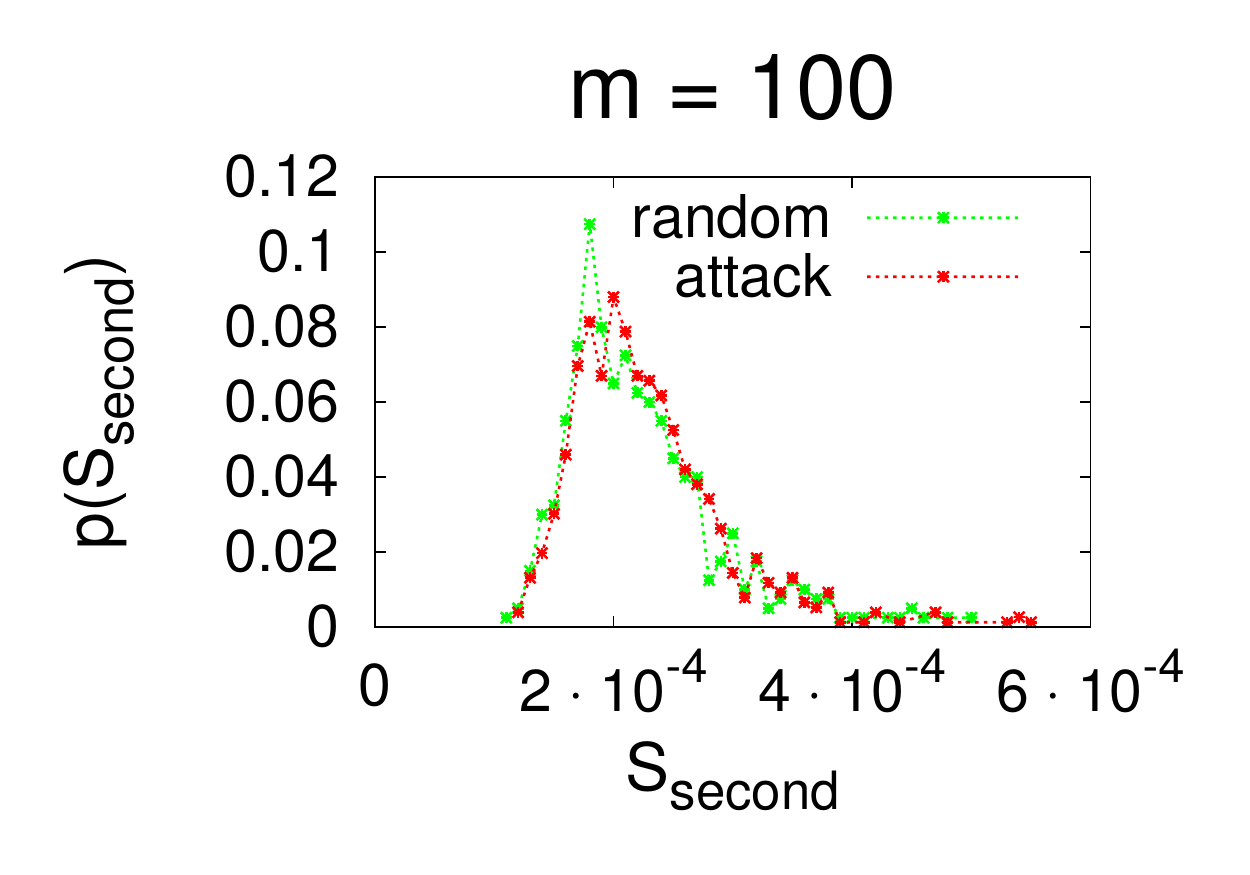}}
\caption{\label{fig:second_total} \textbf{Size of the second largest cluster at S=0.1.} Distribution of the size of the second largest cluster at $S=0.1$ for $\alpha=10$. The size of the second largest cluster is significantly larger (more than two orders of magnitude) in the case of attack for $m=4<m^*$, in agreement with the abrupt first order transition seen in Fig.~2. As mentioned in the main text, this is caused by large modules (i.e. modules that were not much damaged by the attack) that ``dropped'' from the giant component. For $m=6>m^*$, the attack still results in larger second clusters, in comparison to the random attack, because it contains subgraphs (i.e. modules) with more complete internal structure. As $m$ increases, the difference in sizes disappears.}
\end{center}
\end{figure}

\begin{figure}[h]
\begin{center}
\subfigure{\includegraphics[scale=0.42]{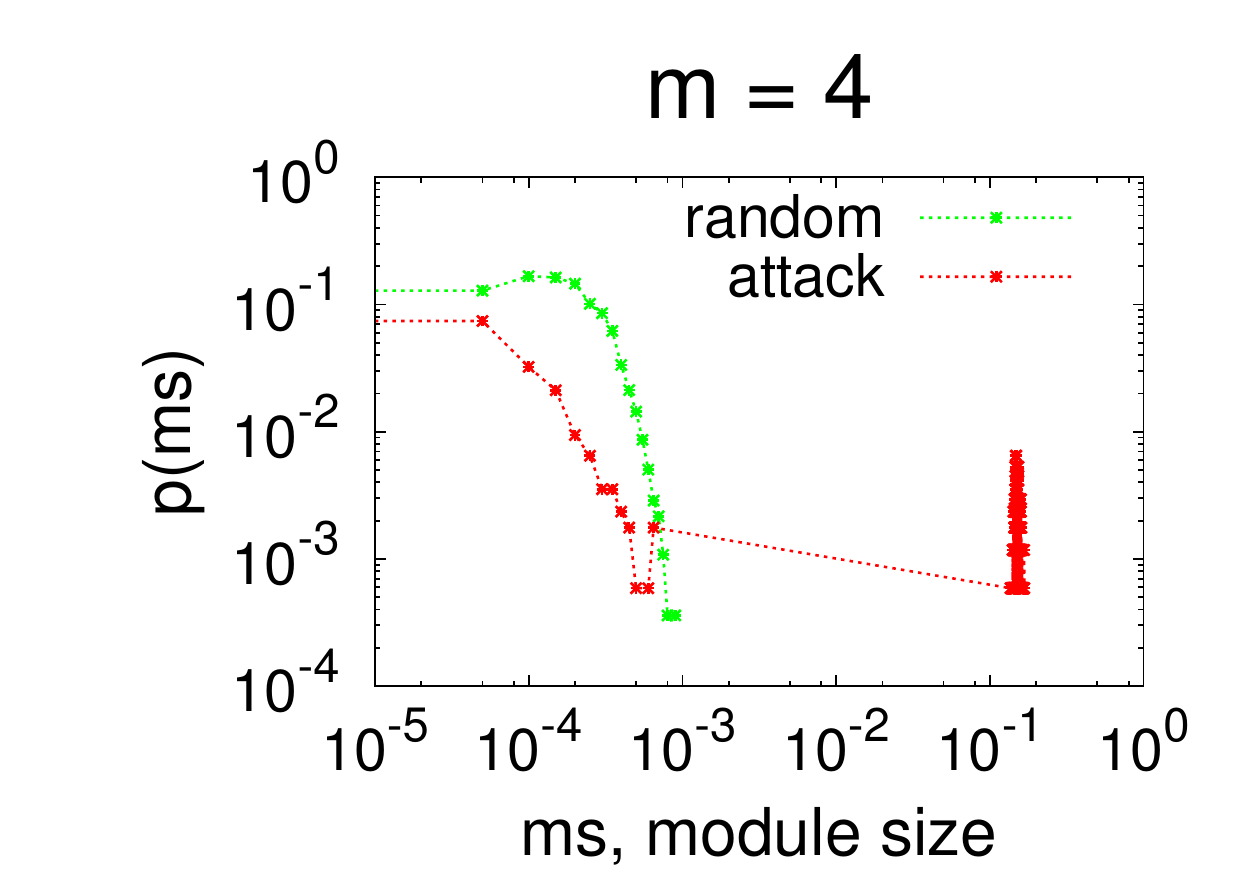}}
\subfigure{\includegraphics[scale=0.42]{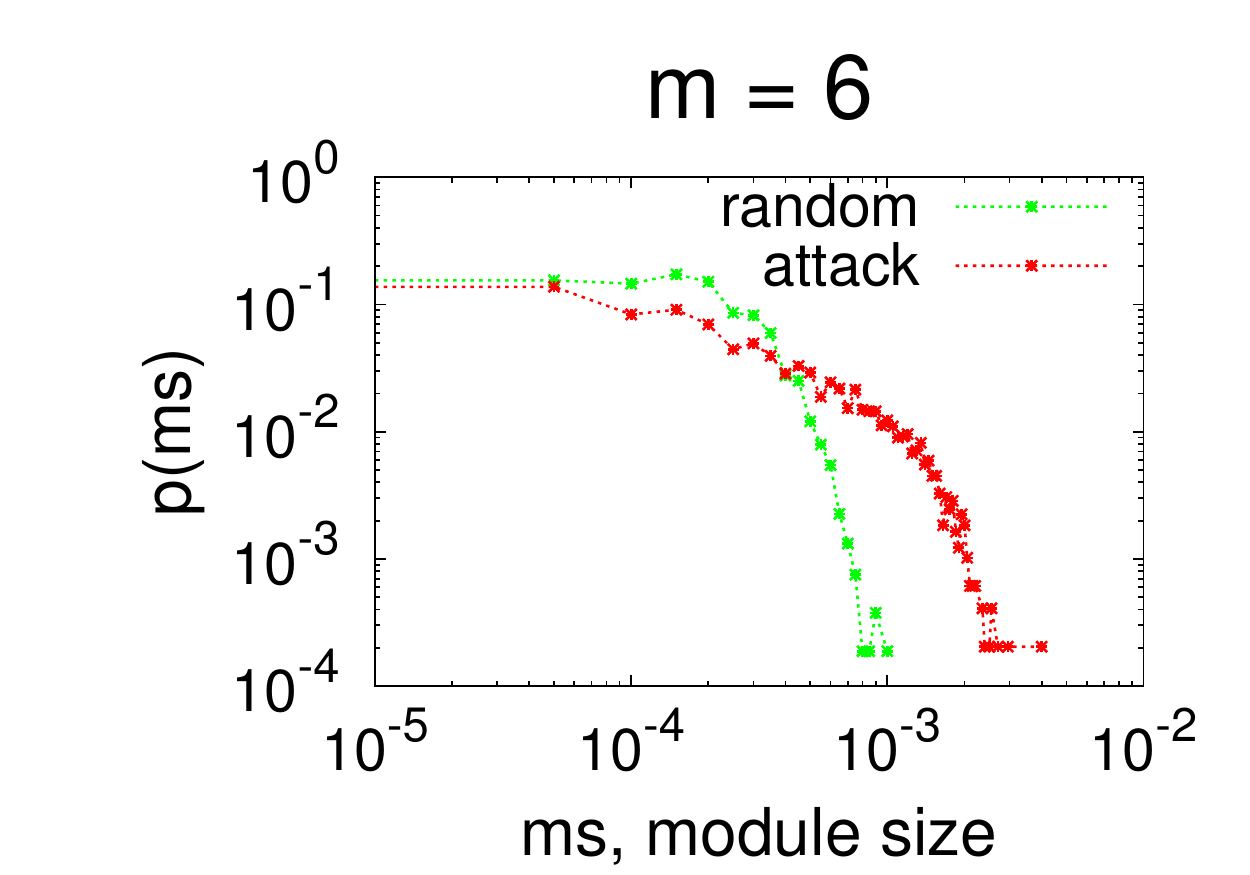}}
\subfigure{\includegraphics[scale=0.42]{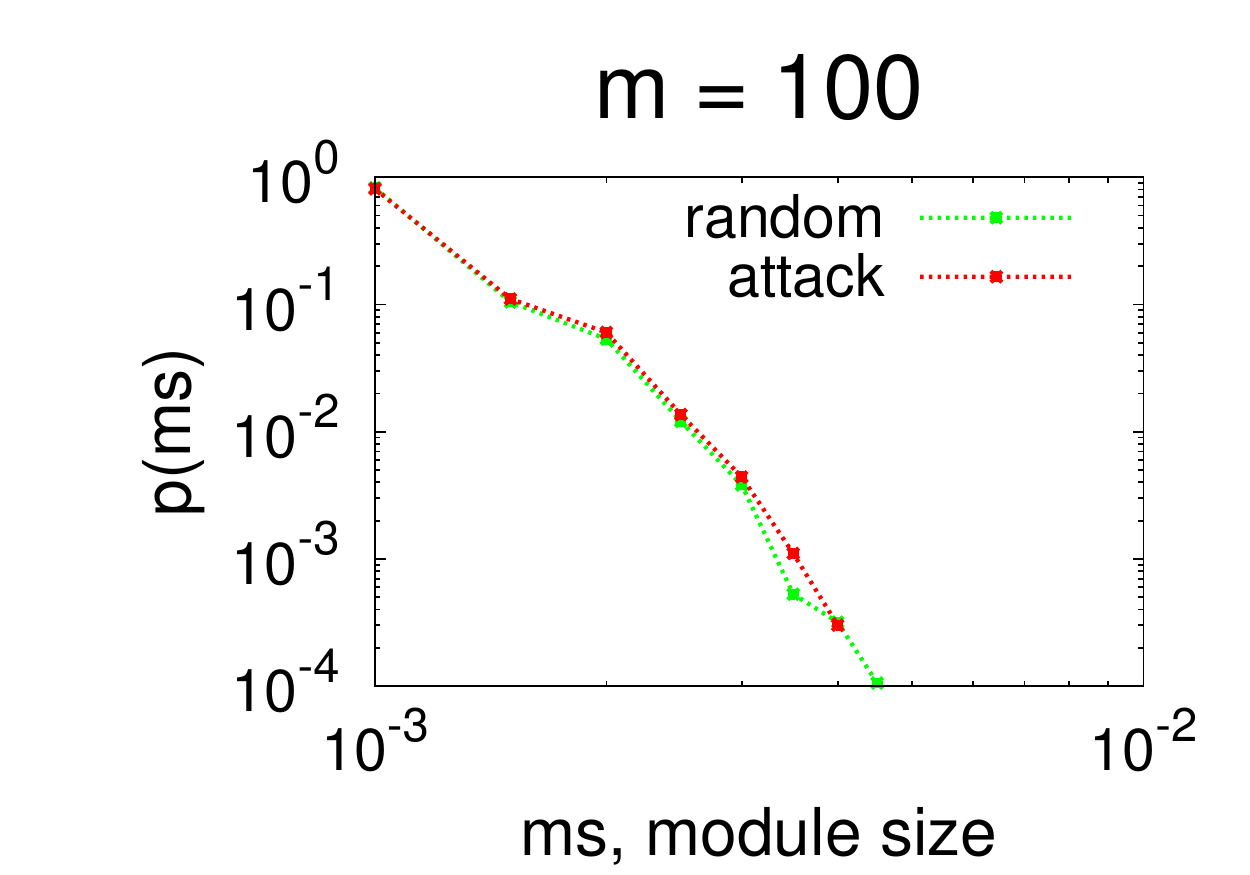}}
\caption{\label{fig:second_cluster_component_size} \textbf{Size of modules in the second largest cluster at S=0.1.} Distribution of the sizes of modules in the second largest component at $S=0.1$ for $\alpha=10$. Modules sizes are normalized by the initial module size. Here we can see that the differences in the size of the second cluster discussed above (Fig.~S5), are indeed originate at big modules that ``dropped'' from the largest component. These modules are getting smaller as $m$ increases since for a large number of modules, the number of interconnected nodes is large, and removing them is damaging the internal structure of modules, just like random node failure.}
\end{center}
\end{figure}

\clearpage 

\section{Betweenness centrality of interconnected nodes in modular structures}
\label{sec:bc}

\begin{figure}[h]
\begin{center}
\subfigure{\includegraphics[scale=0.37]{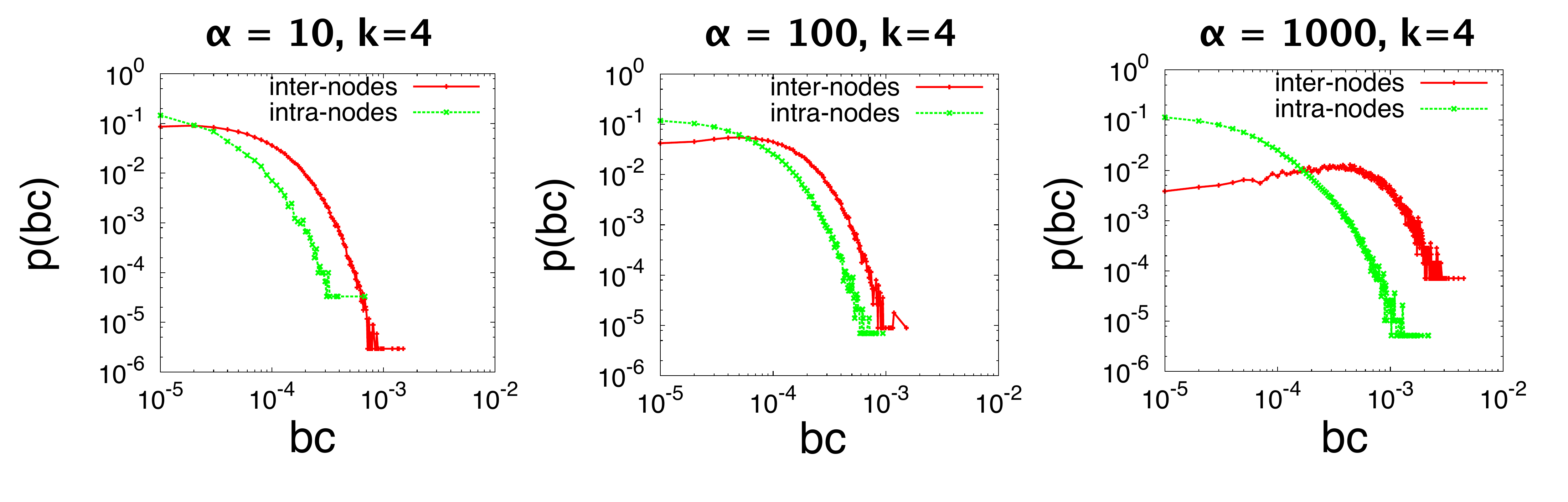}}
\caption{\label{fig:bet} {\bf Betweenness centrality of interconnected nodes for various \boldmath{$\alpha$}.} Betweenness centrality of inter-nodes (nodes that have at least one interconnection) and intra-nodes (nodes with only intraconnections) in networks of size $N=100\thinspace000$ with $m=10$ modules, mean degree $k=4$ and $\alpha=10,100,1000$ respectively. This figure further illustrates the point that in modular structures the interconnected nodes have high betweenness centrality.}
\end{center}
\end{figure}

\end{document}